\documentclass[fleqn,usenatbib]{mnras} 

\usepackage[T1]{fontenc}
\usepackage{newtxtext,newtxmath}

\usepackage{graphicx}
\usepackage{epsfig}
\usepackage{float}
\usepackage{caption}
\usepackage{subfigure}
\usepackage[figuresright]{rotating}

\usepackage{amsmath}

\usepackage{amssymb}
\usepackage{amsfonts}
\usepackage{bm}
\usepackage[mathscr]{euscript}
\usepackage{units}

\usepackage{longtable}
\usepackage{booktabs}
\usepackage{multirow}
\usepackage{threeparttable}

\usepackage{algorithm}
\usepackage{algpseudocode}
\usepackage{listings}

\usepackage{tikz}
\usetikzlibrary{shapes.geometric, arrows.meta, positioning, calc}

\usepackage{color}
\usepackage{soul}

\usepackage{lineno}

\DeclareRobustCommand{\VAN}[3]{#2}
\let\VANthebibliography\thebibliography
\def\thebibliography{\DeclareRobustCommand{\VAN}[3]{##3}\VANthebibliography}

\newcommand\be{\begin{eqnarray}}
\newcommand\ee{\end{eqnarray}}

\title[Real-Color Survey of the Coma Field with the Mephisto]{Simultaneous Real-Color Time-Domain Survey of the Coma Field with the Multi-channel Photometric Survey Telescope (Mephisto)}

\author[Pan et al.]{
Yu Pan$^{1}$,
Yuan-Pei Yang$^{1}$\thanks{E-mail: ypyang@ynu.edu.cn (YPY)},
Ye-Hao Cheng$^{1}$,
Brajesh Kumar$^{1}$\thanks{E-mail: brajesh@ynu.edu.cn (BK)},
Jie-Wei Zhao$^{1}$,
Yongzhi Cai$^{2,3}$,\newauthor
Xinlei Chen$^{1}$,
Jian Cui$^{1}$,
Guowang Du$^{1}$,
Yuan Fang$^{1}$,
Helong Guo$^{4,5}$,
Panpan Huang$^{1}$,\newauthor
Kaushik Chatterjee$^{1}$,
Ziwei Li$^{1}$,
Jianhui Lian$^{1}$,
Chenxu Liu$^{1}$,
Xiangkun Liu$^{1}$,
Chenxi Shang$^{1}$,\newauthor
Tao Wang$^{1}$,
Chunsheng Xiao$^{1}$,
Jinghua Zhang$^{1}$,
Kong-Jun Zhang$^{1}$,
Shiyan Zhong$^{1}$,\newauthor
Zihao Zhou$^{1}$,
Dan Zhu$^{1}$,
Xufeng Zhu$^{1}$,
Xingzhu Zou$^{1}$,
and Xiaowei Liu$^{1}$\thanks{E-mail: x.liu@ynu.edu.cn (XWL)}
\\
$^{1}$South-Western Institute for Astronomy Research, Yunnan Key Laboratory of Survey Science, Yunnan University, Kunming, Yunnan 650504,\\ People's Republic of China\\
$^{2}$INAF - Osservatorio Astronomico di Padova, vicolo dell’Osservatorio 5, I-35122 Padova, Italy\\
$^{3}$International Centre of Supernovae (ICESUN), Yunnan Key Laboratory of Supernova Research, Yunnan Observatories, Chinese Academy \\ of Sciences (CAS),  Kunming, 650216, China\\
$^{4}$Research Center of Astronomy, QingHai University, Xining, 810016, People's Republic of China\\
$^{5}$Department of Physics and Astronomy, QingHai University, Xining, 810016, People's Republic of China\\
}

\date{Accepted XXX. Received YYY; in original form ZZZ}

\pubyear{2026}

\begin{document}

\label{firstpage}
\pagerange{\pageref{firstpage}--\pageref{lastpage}}
\maketitle

\begin{abstract}

We present the results from a high-cadence, multi-channel time-domain survey targeting the Coma galaxy cluster, utilizing the newly commissioned 1.6-m Multi-channel Photometric Survey Telescope (Mephisto). Mephisto's pioneering mosaic focal-plane design enables strictly simultaneous imaging in three distinct optical bands ($u/v, g/r, i/z$). This unique capability allows for the extraction of true, instantaneous spectral energy distributions (SEDs) of fast-evolving transients. Our survey continuously monitors a 20-square-degree footprint centered on the Coma cluster. By strategically targeting this massive local structure, we exploit cosmic anisotropy to significantly amplify the volumetric discovery rate of stellar explosions. During the initial six-month pilot operation, we successfully detected and tracked nine new supernovae (SNe). Remarkably, our high-cadence, multi-band strategy allowed us to capture ultra-early rising light curves and pre-discovery signals (e.g., SN 2026dyr) hours to days prior to formal discovery alerts. These instantaneous early-phase colors allow us to reconstruct intrinsic SEDs without temporal interpolation, providing reliable constraints on the physical parameters of early supernova (SN) evolution. Furthermore, our observed supernova yield shows good agreement with theoretical forecasts based on local stellar mass and star-formation rate overdensities. These pilot results validate the efficiency of the Mephisto Coma survey, demonstrating its objective capability to systematically explore the local transient population and provide rapid optical localizations for future multi-messenger events.

\end{abstract}


\section{Introduction} 

The exploration of the dynamic universe is currently undergoing a revolution, yet it remains fundamentally constrained by the observational limitations of traditional facilities. Specifically, the inevitable time delay introduced by sequential filter switching smears out the intrinsic color evolution of rapidly varying transients. To accurately decipher the extreme physics governing the nascent stages of stellar explosions and to identify fast-evolving anomalous sources, obtaining simultaneous multi-wavelength coverage is imperative. To address this critical observational bottleneck, the newly commissioned 1.6-m Multi-channel Photometric Survey Telescope (Mephisto), developed by Yunnan University, offers a new approach to wide-field time-domain observations.

The Mephisto telescope is a 1.6-m wide-field, multi-channel photometric survey facility operated by the South-Western Institute of Astronomy Research (SWIFAR) at Yunnan University and located at the Lijiang Observatory \citep[IAU code: 044,][]{Liu2019,Yuan2020}. It is designed as a general-purpose time-domain survey instrument optimized for high-cadence, multi-band imaging. The optical system covers six photometric bands ($u$, $v$, $g$, $r$, $i$, and $z$) over a continuous wavelength range of 320--1050 nm, enabling multi-band monitoring of time-domain variability(X.~K.~Liu et al.\ 2026). Mephisto can perform simultaneous three-channel imaging using either the $ugi$ or $vrz$ filter combination.

Before October 2025, Mephisto operated in the single-CCD configuration. In this configuration, the $uv$ and $gr$ channels were equipped with Andor single-chip CCDs with a pixel scale of $0.429''$/pixel, while the $iz$ channel used a detector developed by the National Astronomical Observatories of China (NAOC) with a pixel scale of $0.286''$/pixel. Each channel had a field of view (FoV) of approximately $43\arcmin\times43\arcmin$. The system achieved first light in October 2022 and demonstrated full three-channel operation in December 2023. In 2024, the single-CCD configuration was used for survey observations covering more than 1000 square degrees of sky. These observations demonstrated the feasibility of simultaneous multi-band time-domain imaging and provided observational experience for the subsequent upgrade to the mosaic-camera configuration(X.~K.~Liu et al.\ 2026).

Since October 2025, Mephisto has operated in the mosaic-camera configuration, which adopts a mosaic focal-plane design to enable wide-field time-domain surveys. In this configuration, all three channels employ CCD detectors developed by the NAOC and share a unified focal-plane architecture, resulting in a consistent pixel scale of $0.286''$/pixel across all three channels. The current FoV is approximately $2~\mathrm{deg}^2$. The unified detector architecture provides consistent imaging characteristics across the three channels and facilitates photometric calibration and astrometric alignment, which are important for differential imaging and multi-band variability analysis. Compared with the single-CCD configuration, the mosaic-camera configuration provides a larger FoV and a more uniform detector setup, which can improve the consistency of multi-band light-curve reconstruction. This configuration is particularly well suited for high-cadence, wide-field surveys of nearby galaxies and transient phenomena \citep{Cai2026A&A...707A.157C}.

Taking advantage of Mephisto's wide FoV, we selected regions with relatively high galaxy densities as the targets of our nearby-galaxy time-domain survey. The deployment of such a facility is exceptionally well-suited for probing the local universe ($D \lesssim 200$\,Mpc), which serves as an irreplaceable astrophysical laboratory. Extensive surveys of the local volume are urgently needed to bridge the well-documented ``luminosity desert'' between the brightest classical novae ($M_V \sim -10$\,mag) and sub-luminous SNe ($M_V \sim -16$\,mag), a phase space likely populated by peculiar intermediate-luminosity optical transients \citep{Cai2019,Cai2022}. Furthermore, the nascent fields of multi-messenger astronomy, driven by gravitational wave interferometers, high-energy neutrino observatories, and cosmic ray detectors, are inherently restricted to the local universe due to instrumental sensitivity limits and physical attenuation effects, such as the Greisen-Zatsepin-Kuzmin (GZK) effect. Because these multi-messenger facilities often yield coarse spatial localizations, a wide-field, high-cadence optical survey is crucial for identifying precise optical counterparts, thereby determining the host galaxies and progenitor environments of these extreme events.

On the other hand, high-cadence surveys of nearby galaxies provide a powerful means of discovering and characterizing the early evolution of SNe. Capturing the extremely early epochs of stellar explosions is critical because the initial emission carries uncontaminated signatures of the progenitor star and information about the immediate surroundings, which are quickly erased as the ejecta expands and cools. For core-collapse supernovae (CCSNe), early optical colors can reveal crucial information about excessive UV emission resulting from the violent collision between the fastest supernova ejecta and dense, unshocked circumstellar material (CSM). Beyond core-collapse events, ultra-early multi-band observations are equally pivotal for Type Ia supernovae. High-cadence early color data can reveal the presence of a non-degenerate companion star by detecting the subtle, brief blue bump \citep[e.g.][]{Kasen-2010K} caused by the supernova ejecta crashing into the companion, or differentiate between explosion mechanisms (e.g., double-detonation models versus standard Chandrasekhar-mass explosions). Finally, simultaneous multi-band capabilities are perfectly tuned to discover and characterize newly recognized classes of extreme explosions, such as fast blue optical transients (FBOTs), which evolve too rapidly to be thoroughly studied by conventional surveys.

To maximize the transient yield within this local volume, we have designed a dedicated observational footprint covering 20 square degrees centered on the Coma galaxy cluster, as shown in Figure \ref{coma_survey_field}. The Coma cluster is one of the nearest and richest galaxy clusters in the local universe, located at a redshift of $z\simeq0.023$ and a distance of approximately 100~Mpc. The virial radius of Coma is estimated to be $\sim2.9$~Mpc \citep{Lokas03}. The cluster contains more than a thousand member galaxies and has a total mass of order $10^{15}~M_\odot$, making it one of the most massive gravitationally bound structures in the nearby universe \citep[e.g.,][]{Kubo07}. Its central region is dominated by two giant elliptical galaxies, NGC 4874 and NGC 4889, while the broader Coma supercluster contains numerous galaxy groups and filamentary structures extending into the surrounding large-scale environment \citep[e.g.,][]{Carter08}. The substantial angular extent of the cluster and its surrounding structures makes a wide-field survey particularly advantageous for capturing a large fraction of its galaxy population within a single coherent observational footprint. The high galaxy density and diverse range of galactic environments within and around the cluster make the Coma region an especially valuable laboratory for studying galaxy evolution and transient phenomena. In particular, its rich population of potential transient host galaxies provides a favorable target for wide-field time-domain surveys aimed at maximizing the discovery rate of rare and rapidly evolving optical transients.

By targeting such a massive, overdense region, we can naturally exploit cosmic anisotropy to dramatically boost the volumetric rate of transient events compared to blind-field surveys \citep[see][for details]{Cheng26}. Our observational strategy tessellates this footprint into 10 distinct Mephisto pointings. Utilizing a highly efficient minute-scale operational cycle per pointing, the entire 20-square-degree region can be mapped in one filter triad in just 10 minutes. By alternating between the $u-g-i$ and $v-r-z$ configurations, we can achieve an ultra-rapid localized scanning cadence of $\sim 10$ minutes between the two filter sets, while ensuring that the inter-night cadence across all six optical bands is $\leq 1$ day. 
This combination of simultaneous multi-color imaging, high-cadence monitoring, and a densely populated galactic environment provides a unique observational opportunity to systematically explore rare and rapidly evolving optical transients. 

The paper is organized as follows. In Section~\ref{sec:pipeline}, we detail the observational strategy, data reduction pipeline, and transient detection methodology of the Mephisto Coma field survey. In Section~\ref{sec:results}, we present the initial pilot survey results, highlighting the discovery and real-time multi-color tracking of nine new supernovae. In Section~\ref{Meph-SN-disc}, we delve into the early-phase physics of selected Type Ia and Type IIP supernovae, deriving physical constraints from their rapid photometric evolution.  In Section~\ref{sec:event_rate}, we establish our theoretical expectations for transient event rates based on local cosmic overdensities and quantitatively evaluate our empirical yield, extending the forecasts to rare multi-messenger sources. In Section~\ref{sec:broad_science}, we explore the broader scientific potential of the survey for discovering other extreme transients, such as tidal disruption events and intra-cluster supernovae. Finally, the results and future prospects are discussed and summarized in Section~\ref{sec:conclusion}.

\section{Survey Performance and Data Processing} 
\label{sec:pipeline}

\subsection{Coma Field Survey Strategy and Data Reduction Pipeline}

The Coma field survey is a dedicated time-domain observing program conducted with the Mephisto telescope, aiming to monitor nearby galaxies through systematic multi-epoch, multi-band imaging. The survey covers a total area of $22.58~\mathrm{deg}^2$, with an observing strategy designed to balance temporal cadence and spatial coverage, and further optimized following the upgrade to the mosaic-camera system.

The observations cover six photometric bands, $u$, $g$, $i$, $v$, $r$, and $z$, organized into two three-band filter sets, $ugi$ and $vrz$. Each exposure has an integration time of 45~s, and the mean seeing across the six bands is approximately $1.9\arcsec$.

\begin{figure}
  \centering
  \includegraphics[width = 1.0\linewidth, trim = 0 0 0 0, clip]{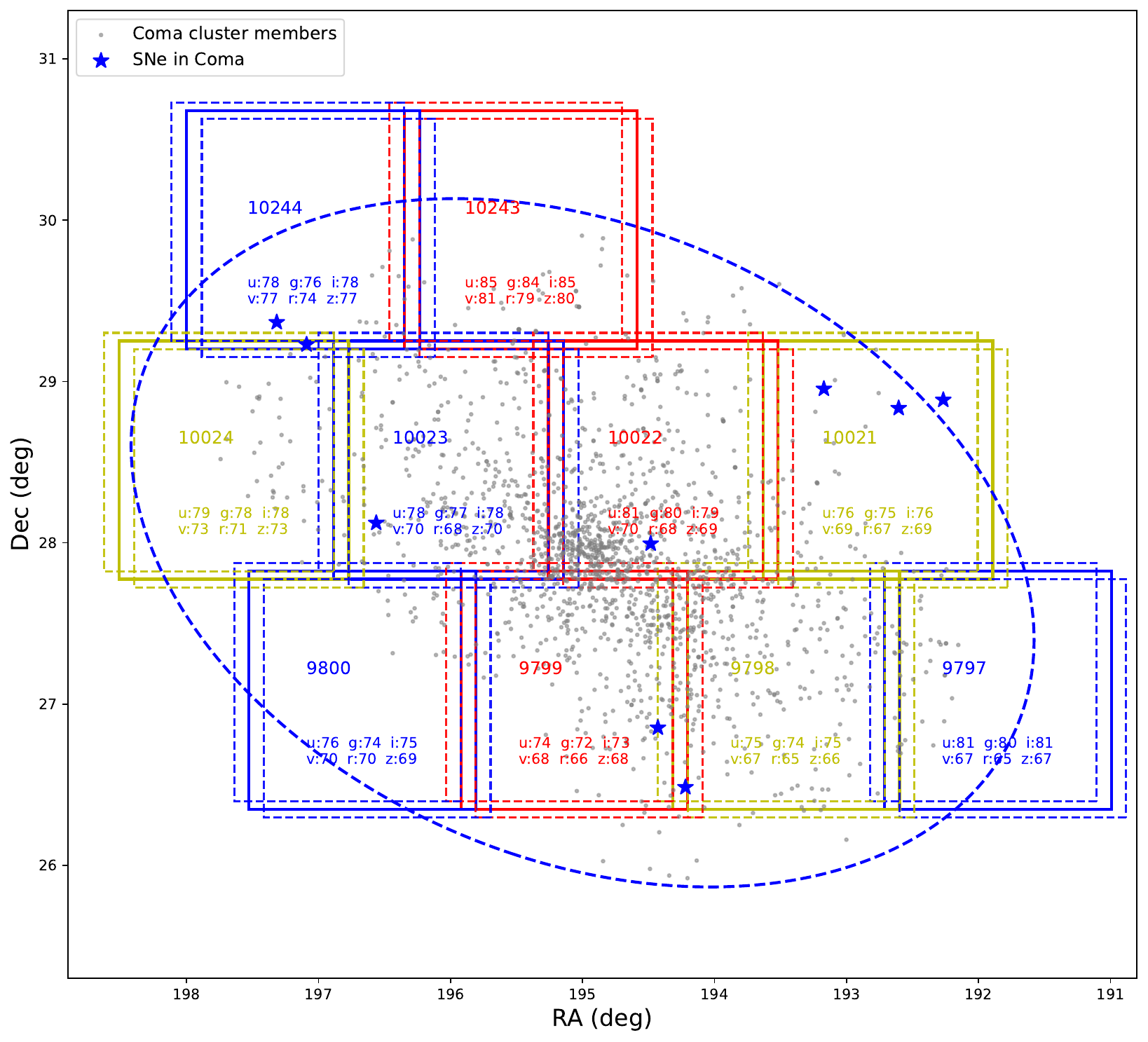}
  \caption{
Survey footprint of the Coma field. The Mephisto survey divides the Coma field into 10 primary survey fields (solid outlines) and 20 dithered fields (dashed outlines). The number of observations in each photometric band for each field is indicated on the map. The blue stars mark the positions of the nine supernovae identified in the Coma field. 
  }
  \label{coma_survey_field}
\end{figure}

The Coma field consists of 10 primary survey fields ($m_1$--$m_{10}$). To compensate for the gaps between CCDs in the mosaic focal-plane design, each primary field is accompanied by two additional dither fields ($d1$ and $d2$), resulting in a total of 30 observing fields. The spatial distribution of these fields is shown in Figure~\ref{coma_survey_field}, where the primary fields are outlined by solid lines and the corresponding dither fields by dashed lines.

During the early phase of the survey, one filter set, either $ugi$ or $vrz$, was typically observed each night, covering all three groups of fields, including the primary fields ($m$), the first dither fields ($d1$), and the second dither fields ($d2$). This observing strategy enabled complete coverage of the survey footprint within a single night, but resulted in relatively sparse temporal sampling of individual fields. 

Since December 23, 2025, the observing strategy has been revised to improve the temporal cadence. Both $ugi$ and $vrz$ filter sets are observed within the same night, while only one of the three field groups is observed per night, namely the primary fields ($m$), the first dither fields ($d1$), or the second dither fields ($d2$). This observing mode retains short-timescale multi-band coverage while increasing the revisit frequency of individual fields, thereby providing improved temporal sampling for the detection and characterization of variable and transient sources in nearby galaxies.

All raw images are processed through standard CCD reduction procedures, including bias subtraction, dark-current correction, and flat-field correction, to remove instrumental signatures prior to scientific analysis(Y.~Fang et al.\ 2026).
Photometric homogenization is then performed across CCDs and readout amplifiers by transforming the instrumental zeropoints to a common reference system. This procedure ensures consistent flux calibration across the mosaic focal plane and provides a uniform photometric baseline for subsequent image combination and difference imaging(J.~H.~Zhang et al.\ 2026).

\subsection{Transient Detection Pipeline}

To systematically identify transient and variable sources from the reduced data, we developed a dedicated difference imaging pipeline. The primary objectives of the pipeline are rigorous image quality filtering, the construction of deep static-sky reference templates, and precise PSF-matched image subtraction, thereby producing a robust initial catalog of transient candidates.

Each exposure image is first evaluated using image-quality parameters, including the full width at half maximum (FWHM), ellipticity, background level, and background RMS. Images satisfying the quality criteria of FWHM $<1.8''$, ellipticity $<0.15$, and background and background RMS values within $1\sigma$ of the median values of the full exposure sample are selected for reference image construction. For each target field, filter, and observing epoch, the selected high-quality images are combined to construct deep reference images. During reference image construction, known moving objects are identified to avoid contamination from Solar System objects. The positions of known minor planets are obtained through ephemeris-based positional queries with a matching radius of $3''$, and adaptive masks are applied around their predicted positions according to the image FWHM. Considering that the Mephisto mosaic camera consists of multiple CCDs and independent readout regions (gates), photometric normalization is performed separately for each gate. The zeropoints of individual gates are calibrated to a common value of 25 mag, and the measured fluxes are normalized to an equivalent exposure time of 45 s.

Before image coaddition, background correction is performed independently for each gate. Source masks are first generated for the images, after which the residual large-scale background variations are modeled using a fourth-order Zernike polynomial surface. The background-corrected and photometrically normalized images are then resampled onto a common astrometric grid using SWarp \citep{Bertin2010}. The aligned images are subsequently combined to produce deep reference images for individual CCD chips. These reference images provide high signal-to-noise representations of the static sky and serve as templates for subsequent difference imaging.

Image subtraction is performed using HOTPANTS \citep{Becker2015} to obtain difference images and identify variable and transient sources. Before subtraction, the science images are processed using the same photometric normalization procedure as the reference images. To reduce contamination from satellite trails, cosmic rays, and other non-astronomical artifacts, satellite trails are first identified in the original images using a machine-learning-based detection model, while cosmic rays and other anomalous pixels are identified using sigma clipping. Corresponding weight maps are then generated and used in the subsequent subtraction and source-detection procedures.

Because of the spatial variation of the point spread function (PSF) across the large field of view, each CCD is divided into four quadrants, which are processed independently. For each quadrant, the FWHM values of the science and reference images are first measured. The image with better seeing is then convolved to match the PSF of the image with poorer seeing. Image subtraction is subsequently performed on the PSF-matched images. After removing the overlapping boundary regions between adjacent quadrants, the four quadrants are reconstructed into a complete CCD difference image. Source detection is then performed on the difference images using SExtractor \citep{Bertin1996} to construct the initial transient-candidate catalog.

Transient candidates are subsequently associated across bands by identifying simultaneously observed images in different filters, forming multi-band observation sets such as $ugi$ and $vrz$. For three-band observations, spatial cross-matching is performed using a matching radius of $1''$. Three types of joint catalogs are constructed: sources detected in all three bands, sources detected in at least two bands, and the full union catalog containing all detections. All transient candidates are further cross-matched with external catalogs, including Gaia DR3 \citep{Gaia2023} and SDSS DR16 \citep{Ahumada2020}, to identify known stars, quasars, and galaxies. Known minor planets are additionally queried using ephemeris information and flagged for subsequent candidate classification and quality assessment.

Multi-band association significantly reduces false detections caused by image subtraction artifacts, reducing the initial candidate sample from hundreds to thousands of single-band detections to a manageable sample of several tens of candidates. For candidates detected in two or more bands, $1'\times1'$ image cutouts are extracted from the science, reference, and difference images to construct images for visual inspection (see Figure \ref{fig:panel}). Each panel consists of three rows corresponding to the blue, yellow, and red channels, together with an RGB composite image. From left to right, the panels display the science, reference, and difference images, respectively, allowing direct comparison of the candidates' multi-band morphology, variability signatures, and subtraction residuals. The RGB composite images are generated using STIFF \citep{Bertin2012} to facilitate visual examination of the source morphology and color properties.

\section{Pilot Survey Results: Discovery of Nine Supernovae}
\label{sec:results} 

\begin{figure}
    \centering
    \includegraphics[width = 1.0\linewidth, trim = 0 0 0 0, clip]{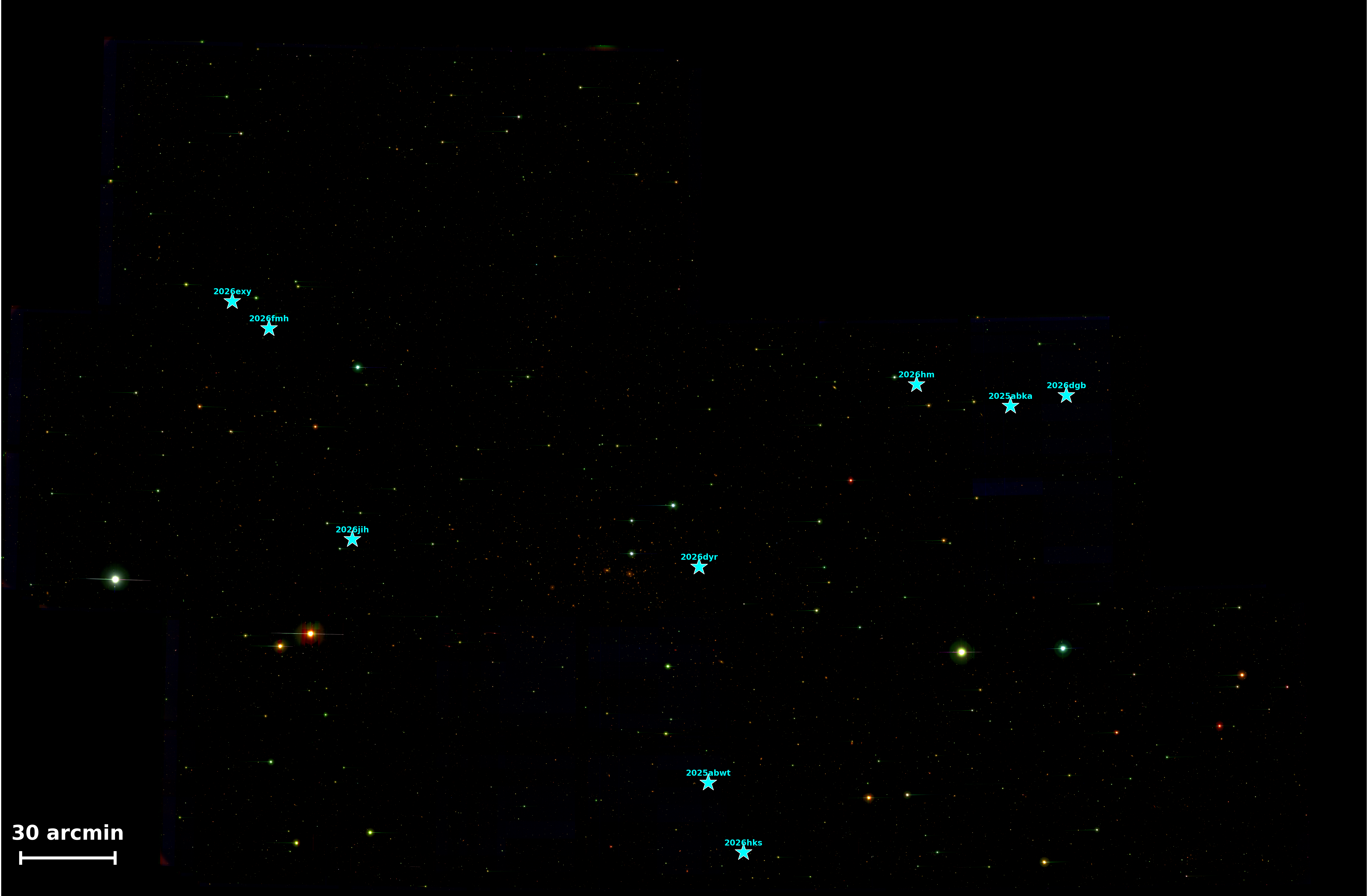}
    \includegraphics[width = 1.0\linewidth, trim = 0 0 0 0, clip]{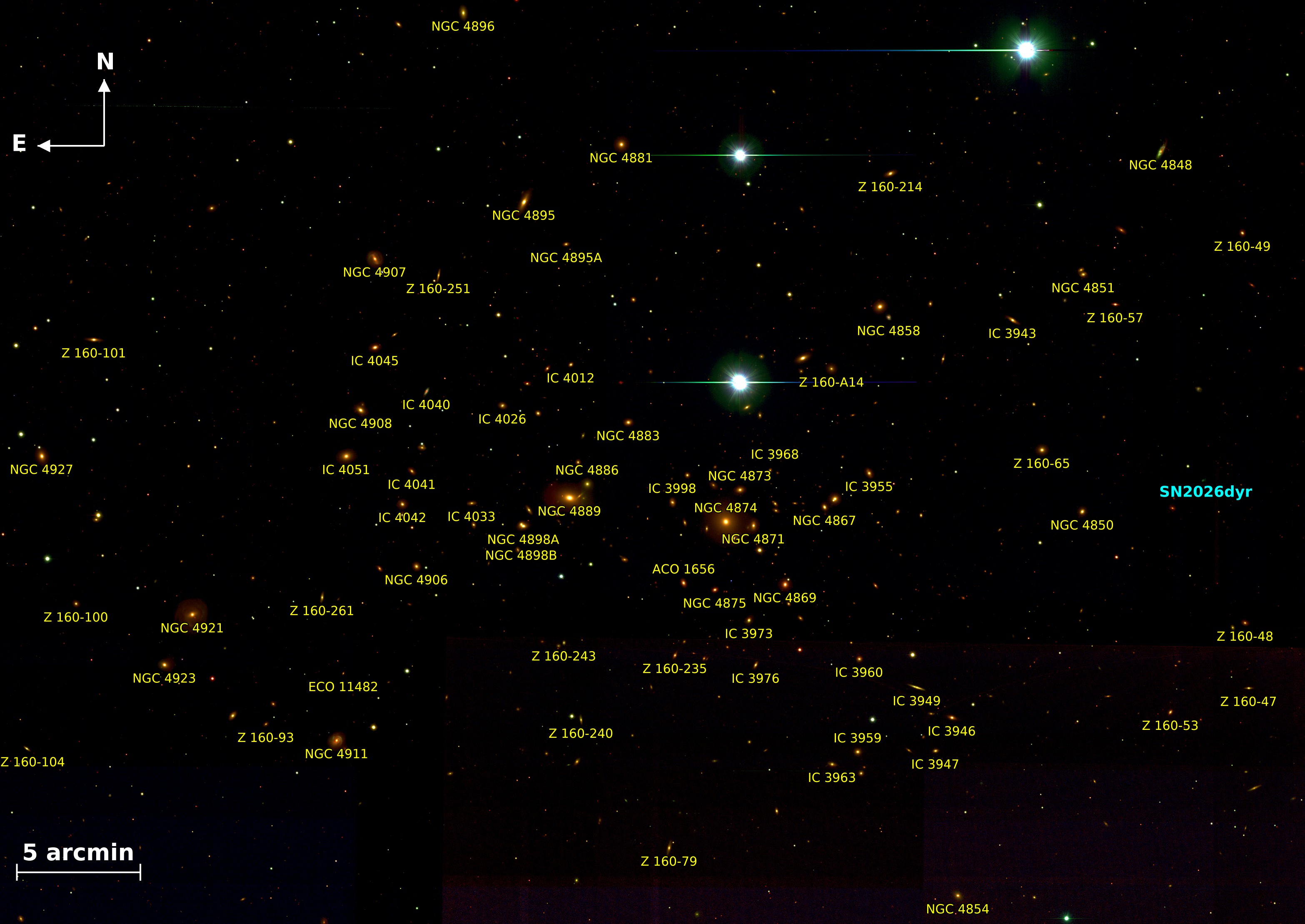}
    \caption{Top panel: location of nine SNe identified in the Coma field survey with the Mephisto telescope. North is up, and East is on the left of the image. Different SNe are indicated with a square symbol. Bottom panel: the $ugi$ color composite image of the central Coma region, observed with Mephisto.}
    \label{nineSNe}
\end{figure}

During the first six months of the Mephisto Coma field survey, covering the commissioning period from October 2025 to April 2026, we monitored an area of approximately $20~\mathrm{deg}^2$ based on the mosaic-camera configuration. Based on visual inspection of the diagnostic panels and cross-matching with the Transient Name Server (TNS), we identified and confirmed nine supernovae within the survey fields, including 4 Type Ia, 2 Type Ibc, and 3 Type II SNe. 
This initial sample demonstrates the feasibility of the adopted observing strategy and data-reduction pipeline for time-domain searches for transient sources in the nearby Universe. The spatial distribution of the nine supernovae is shown in the top panel in Figure~\ref{nineSNe}, while the central region of the Coma field is presented in the bottom panel in Figure~\ref{nineSNe}.

\begin{figure*}
\centering
  \includegraphics[width = 0.9\linewidth, trim = 0 0 0 0, clip]{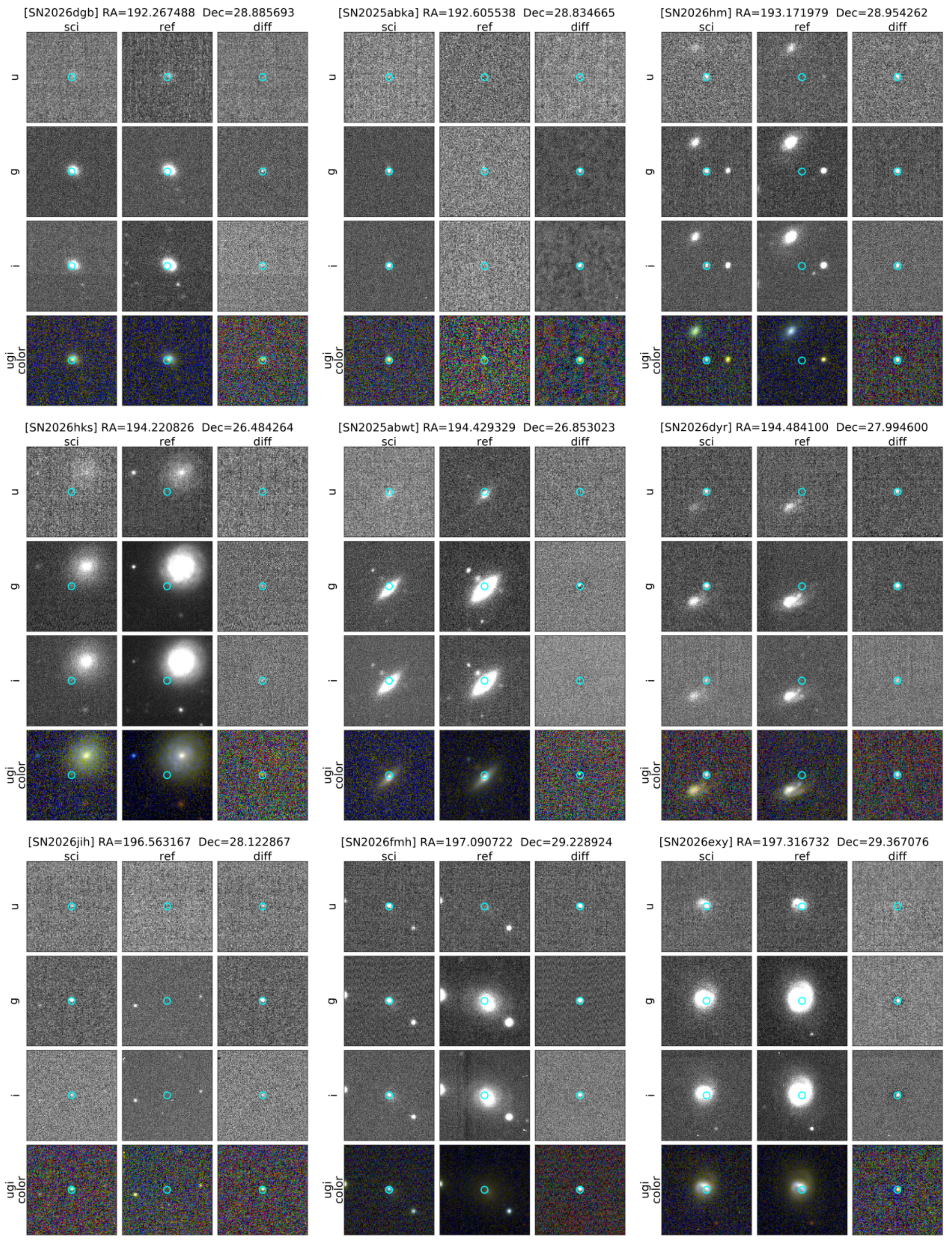}
  \caption{Image cutouts of the identified supernovae. Each stamp covers a $1\arcmin\times1\arcmin$ FoV. For each supernova, the science, reference, and difference images are shown from left to right for a representative epoch, with different rows corresponding to different photometric bands. North is up and east is to the left in each stamp.
} 
  \label{fig:panel}
\end{figure*}

For each supernova, we extracted its light curve using image subtraction, with high-quality images obtained before the discovery date reported by TNS selected to construct the reference template, as shown in Figure~\ref{fig:panel}. For two supernovae that exploded relatively early, historical images obtained during the single-CCD configuration of Mephisto were instead used as the reference templates. The subsequent science images were then subtracted against the corresponding templates to obtain the transient flux at each epoch, and the difference-image panels were regenerated for each source.

Several supernovae show detectable flux excesses within $\sim 1$~day of, or even prior to, the discovery dates reported by the TNS, demonstrating that the high-cadence, multi-band observations of Mephisto can capture early-phase signals following supernova explosions. Their basic properties, cross-matched with the TNS database and ordered chronologically by their reported discovery dates, are summarized in Table~\ref{tab:sne_sample}, with their classifications adopted from the TNS database.
The sample exhibits a diverse range of transient subtypes and progenitor environments. Spectroscopic classifications reveal that the sample consists of both thermonuclear (Type Ia) and massive stellar core-collapse (Types II, IIP, Ib, Ib/c) events. 
Based on their spectroscopic redshifts, two of the discovered transients (SN 2026dgb and SN 2026jih) are background events located well beyond our $150$\,Mpc threshold ($z \simeq 0.061$ and $z \simeq 0.120$, respectively).

\begin{table*}
\centering
\caption{Nine supernovae detected in the Mephisto Coma field survey.}
\label{tab:sne_sample}
\footnotesize
\setlength{\tabcolsep}{5pt}
\begin{tabular}{cccccccc}
\hline
Mephisto Epoch &
R.A. &
Decl. &
TNS Name &
Discovery Date &
Type &
$z$ &
Notes \\
\hline
2026-02-11 & 192.2675 & 28.8857 & 2026dgb  & 2026-02-12 & SN Ia  & 0.061098 & Pre-discovery signals \\
2024-05-13 & 192.6055 & 28.8347 & 2025abka & 2025-10-22 & SN II  & 0.030000 & Host: NGP9 F322-1015955 \\
2026-01-05 & 193.1720 & 28.9543 & 2026hm   & 2026-01-07 & SN II  & 0.032677 & -- \\
2026-03-21 & 194.2208 & 26.4843 & 2026hks  & 2026-03-25 & SN Ib  & 0.025361 & Host: IC 835 \\
2025-07-10 & 194.4293 & 26.8530 & 2025abwt & 2025-10-29 & SN Ia  & 0.021251 & -- \\
2026-02-21 & 194.4841 & 27.9946 & 2026dyr  & 2026-02-22 & SN IIP & 0.015200 & Host: LEDA 83707; pre-discovery signals \\
2026-04-06 & 196.5632 & 28.1229 & 2026jih  & 2026-04-11 & SN Ia  & 0.120000 & Pre-discovery signals \\
2026-03-10 & 197.0907 & 29.2289 & 2026fmh  & 2026-03-13 & SN Ia  & 0.023400 & Host: LEDA 166938; pre-discovery signals \\
2026-03-05 & 197.3167 & 29.3671 & 2026exy  & 2026-03-06 & SN Ib/c & -- & Host: Z 160-151; pre-discovery signals \\
\hline
\end{tabular}

\parbox{\textwidth}{\footnotesize
\textit{Notes.}
Coordinates are given in degrees (J2000).
The ``Mephisto Epoch'' is defined as the epoch of the Mephisto observation closest in time prior to the outburst. Whenever data from the mosaic-camera configuration are unavailable at the relevant epoch, we instead use data from the single-CCD configuration.
Discovery dates are from the Transient Name Server (TNS).
``Pre-discovery signals'' indicates flux variations detected in Mephisto observations before the discovery date.
}
\end{table*}

\begin{figure*} 
\centering
  \includegraphics[width = 1.0\linewidth, trim = 0 0 0 0, clip]{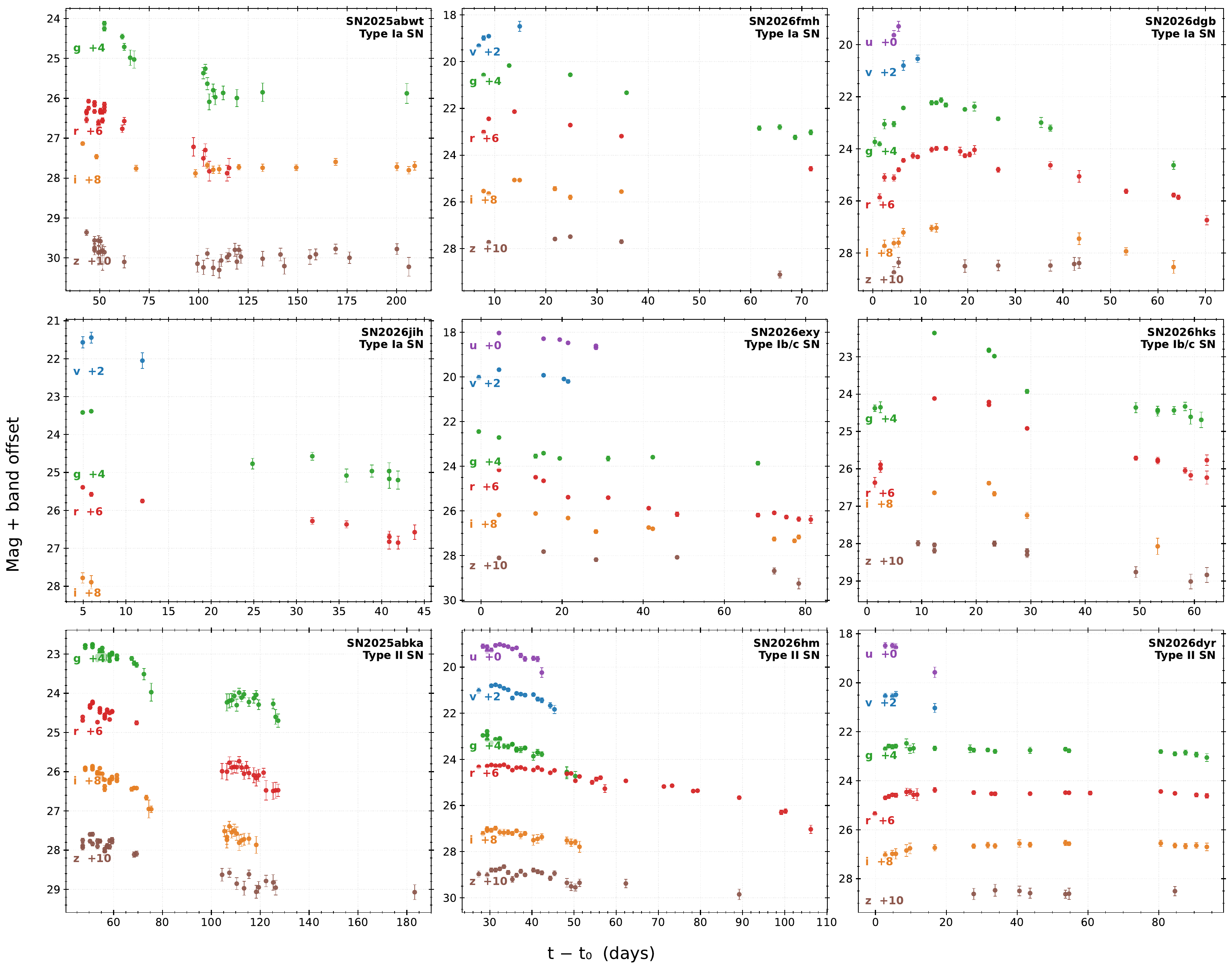} 
   \caption{Multi-band light curves of the nine identified supernovae in the Coma field surveyed by the Mephisto. The classifications are adopted from the Transient Name Server (TNS). The time axis is given relative to the TNS discovery epoch ($t-t_0$). Several objects exhibit detectable pre-discovery emission before $t_0$ (cf. Section~\ref{Meph-SN-disc} and Table~\ref{tab:sne_sample}). The data are available at Science Data Bank: \url{https://doi.org/10.57760/sciencedb.014le}}
  \label{fig:light_curves} 
\end{figure*}

To further illustrate the scientific potential of our survey, Figure~\ref{fig:light_curves} presents the multi-band ($u, v, g, r, i, z$) light curves of all nine supernovae discovered during the pilot phase, together with their host galaxies. The dataset provides dense, high-quality photometric coverage from the near-ultraviolet ($u$-band) to the near-infrared ($z$-band), enabling well-sampled tracking of the photometric evolution of supernovae, particularly during their early phases. 
Because the data points in the blue, yellow, and red channels are recorded strictly simultaneously, the multi-band light curves presented in Figure~\ref{fig:light_curves} are not merely collections of isolated photometric points; rather, every vertical slice through a filter triad represents a true, instantaneous optical SED.

Furthermore, these light curves underscore Mephisto's exceptional sensitivity in the blue and near-ultraviolet regimes ($u$ and $v$ bands). While the initial emission following a core collapse peaks strongly in the UV/blue, these bands are notoriously difficult to monitor continuously from the ground. Our pilot data demonstrate robust, high-signal-to-noise detections in the $u$ and $v$ bands from the very first epochs. These early blue-band data serve as critical anchor points for identifying early shock breakout and early circumstellar material (CSM) interactions, effectively mapping the final, episodic mass-loss history of the massive star prior to its death. 

The exquisite sampling capability of our survey is perfectly illustrated by the specific targets highlighted in our sample. Thanks to our unique 10-minute intra-night cadence, we successfully recorded pre-discovery signals and extremely early rising light curves for several targets, notably SN~2026fmh and SN~2026dgb for the early rising evolution.
In the next section, we will provide a brief description of individual events detected in the Coma field.

\section{Description of individual supernovae in the Coma field}\label{Meph-SN-disc}

\textbf{SN~2025abwt (Type Ia)}:
This SN was discovered (16.72 mag, $r$-ZTF band) on 2025 October 29 by the ZTF group \citep{SN2025abwt}, and the last non-detection was on 2025 October 27 (16.79 mag, $r$-ZTF band). A spectrum obtained with the Palomar 5.1m Hale telescope on 2025 November 30 classified it as a Type Ia event \citep{SN2025abwt-spec}. Mephisto observations of the SN~2025abwt field were obtained in both the single-CCD and mosaic-camera configurations. The field was observed as early as 2025 July 10 with the single-CCD configuration, while observations with the mosaic-camera configuration began relatively late, starting on 2025 December 8, and continued until 2026 May 26.

\textbf{SN~2026dgb (Type Ia)}:
Discovery by ZTF on 2026 February 12 at mag 19.67 in the $g$-ZTF band \citep{2019PASP-ZTF,2026TNSTR-SN2026dgb}. The last non-detection was at 20.05 mag ($r$-ZTF band) on 2026 February 08. Pre-discovery Mephisto observations for this field were taken between 2025 December 8 and 2026 February 11. The field has consecutive night $uvgriz$-band observations between 2026 February 3 and 2026 February 11. It was classified as a Type Ia SN \citep{2026TNSCR.994....1F-2026dgb}.

\textbf{SN~2026fmh (Type Ia)}:
The GOTO survey discovered it on 2026 March 13 at 19.15 mag ($L$-GOTO filter), and the last observation was on 2026 March 12 (limiting magnitude 20.0 mag, $L$-GOTO) \citep{2026TNSTR-2026fmh}. The source was classified as a Type Ia SN using the spectrum taken on 2026 March 15 \citep{2026TNSCR1101....1F}. Pre-discovery Mephisto observations of this field were between 2025 December 8 and 2026 March 10.

\textbf{SN~2026jih (Type Ia)}:
\citet{SN2026jih} discovered SN~2026jih on 2026 April 11 with 0.6m/0.9m Schmidt telescope at Konkoly Observatory, Hungary. The discovery magnitude was 20.07 mag ($w$-P1 band); however, the last non-detection epoch was 2026 March 13. The spectrum taken on 2026 April 13 from the HET-LRS instrument at the Hobby-Eberly Telescope classified it as a Type Ia event before maximum light. Notably, the SN was identified in the Mephisto images ($g,r$ and $i$ band frames) taken on 2026 April 06 and the respective magnitudes were 19.81 $\pm$ 0.13, 19.52 $\pm$ 0.16 and 20.74 $\pm$ 0.20. This identification at an earlier epoch than the reported discovery date confirms that the SN exploded several days earlier than the reported discovery.

\textbf{SN~2026exy (Type Ibc)}:
On 2026 March 06, SN~2026exy was discovered at mag 18.59 ($r$-ZTF band) by the ZTF group \citep{SN2026exy}, and the last non-detection was on 2026 February 28 (20.26 mag, $g$-ZTF band). Here, we note that the SN was clearly identified in the $v,g,r,i,z$ Mephisto frames taken on 2026 March 03 at magnitudes: $v$ = 20.99 $\pm$ 0.36, $g$ = 20.05 $\pm$ 0.11, $r$ = 19.87 $\pm$ 0.12, $i$ = 20.19 $\pm$ 0.15 and $z$ = 20.26 $\pm$ 0.26. This confirms the SN discovery about 3 days earlier than the reported one. The spectrum obtained on 2026 April 6 by the Walter Baade Magellan 6.5-m telescope (LCO) classified it as a Type Ibc event \citep{SN2026exy-spec}.

\textbf{SN~2026hks (Type Ib)}:
\citet{SN2026hks} discovered SN~2026hks on 2026 March 25 using ZTF-Cam at 21.05 mag ($r$-ZTF band). The last non-detection was on 2026 March 24 (20.92 mag, $r$-ZTF band). It was classified as a Type Ib SN from a spectrum taken on 2026 May 9 \citep{SN2026hks-spec}. The pre-discovery Mephisto imaging of the field was from 2025 December 8 to 2026 March 21. Further, the SN field was observed spanning from 2026 March 26 to 2026 May 25. 

\textbf{SN~2025abka (Type II)}:
ZTF discovered this SN on 2025 October 22; \citet{SN2025abka} reported a discovery magnitude of 18.57 in the $r$-ZTF band. The last detection was on 2025 October 21, with a magnitude of 17.88 in the $r$-ZTF band. It was classified as a Type II SN based on a spectrum obtained on 2025 November 13 with the SEDM instrument at the 1.5 m Palomar facility \citep{SN2025abka-2025TNS}. Mephisto observations of the SN~2025abka field were obtained in both the single-CCD and mosaic-camera configurations. The field was observed as early as 2024 May 13 with the single-CCD configuration, while the observations with the mosaic-camera configuration began on 2025 December 8 and continued until 2026 May 26. The resulting temporal sampling is relatively sparse.

\textbf{SN~2026hm (Type II)}:
\citet{2026hm-TNSTR..77C} from Pan-STARRS group discovered SN~2026hm (PS26at) on 2026 January 07 at 18.83 mag ($i$-P1 band). A spectrum from Palomar 5.1m Hale, taken on 2026 January 14, classified it as a Type II SN \citep{2026hm-spec}. The latest pre-discovery Mephisto observations of this field were on 2026 January 5. Between 2026 February 3 and 2026 May 25, Mephisto has multi-band observations at many epochs.

\textbf{SN~2026dyr (Type II)}:
The discovery of SN~2026dyr (coordinates: 12$^{\rm h}$~57$^{\rm m}$~56\fs189, +27\degr ~59\arcmin ~40\farcs50, J2000)  was reported \citep{2026TNSTR-SN2026dyr} on UT 2026 February 22.11 (JD~2461093.61) by the Gravitational-wave Optical Transient Observer survey \citep{GOTO-2022Steeghs} with a discovery magnitude of 19.08 ($L$-GOTO filter). The last non-detection was on UT 2026 February 21.06 (20.1 mag). A spectrum of SN~2026dyr obtained on 2026 February 25 with the Nordic Optical Telescope classifies it as a Type IIP SN \citep{2026TNSTR-SN2026dyr}. A detailed investigation of this SN is presented as follows.

Considering that SN~2026dyr was discovered very early after the explosion (within 1 day) and that well-sampled Mephisto data were available, we analysed its observational properties in detail and also compared it with other well-observed events.
In the Mephisto survey, the first clear detection of SN~2026dyr was on UT 2026 February 21.93 at a magnitude of 19.67$\pm$0.10 ($r$-band), confirming that the SN exploded earlier than the reported date. We consider the explosion epoch of SN~2026dyr to be MJD~$61092.4\pm0.5$ (see the following Sect.~\ref{sec:26dyr:lcprop}), the midpoint of the last non-detection and the first detection of SN in the Mephisto image.

The most likely host galaxy of SN~2026dyr is LEDA~83707, located at $\alpha = 12^{\rm h}57^{\rm m}56\fs796$, $\delta = +27\degr59\arcmin30\farcs30$ (J2000). 
We adopt the host galaxy redshift of $z = 0.015226 \pm 0.000004$, retrieved from the NASA/IPAC Extragalactic Database (NED), for the SN. Assuming a flat $\Lambda$CDM cosmology with $H_0 = 70\,\mathrm{km\,s^{-1}\,Mpc^{-1}}$, $\Omega_{\rm m} = 0.3$, and $\Omega_{\Lambda} = 0.7$ \citep{Freedman2019ApJ...882...34F}, this yields a luminosity distance of $D_{\rm L} = 66.0\,\mathrm{Mpc}$ and a corresponding distance modulus of $\mu = 34.10$~mag. The Galactic extinction towards SN~2026dyr is $E(B-V)_{\rm MW} = 0.0092 \pm 0.0007$~mag, which can be obtained from the dust map of \cite{2011schlafly}. Since no prominent Na~\textsc{i}~D absorption lines are detected in the early-phase spectra, host-galaxy extinction is assumed to be negligible ($E(B-V)_{\rm host} = 0$~mag). Consequently, the total line-of-sight extinction is taken as $E(B-V)_{\rm tot} = 0.009 \pm 0.001$~mag.

During the Mephisto commissioning, this field was monitored (in multi-band) at many epochs between 2025 December 08 and 2026 May 26. The latest Mephisto observations were only a day before the explosion (see Section~\ref{sec:26dyr:lcprop}). We used this information to estimate the explosion epoch of this target as discussed in the following section.

\subsection{Light curve properties of SN~2026dyr}
\label{sec:26dyr:lcprop}
\begin{figure}
\centering
\includegraphics[width=0.9\linewidth]{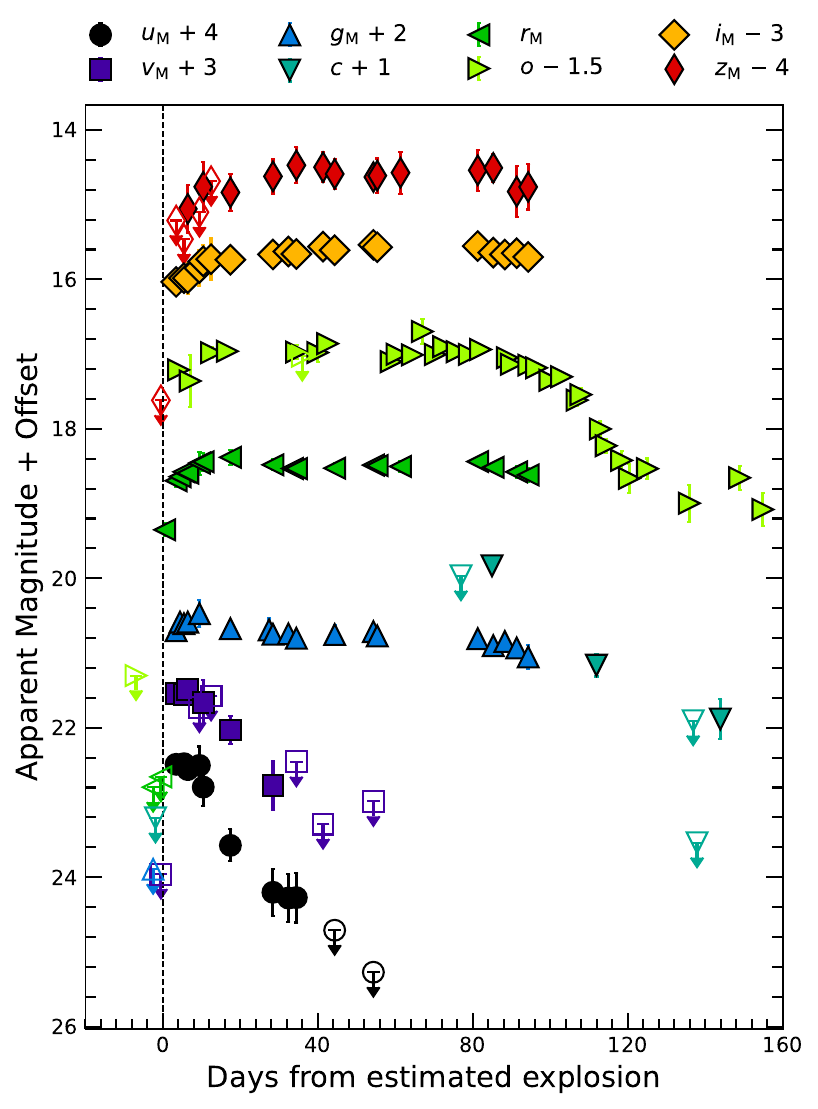}
\caption{Multi-band light curve evolution of SN~2026dyr. Phases are calculated from the estimated explosion epoch. Different bands are presented with different colors and symbols. The indicated offset is applied for clarity. The upper limit points are shown with hollow symbols.}
\label{fig:26dyr:lc}
\end{figure}

The multi-band light curves of SN~2026dyr are presented in Figure~\ref{fig:26dyr:lc}. For comparison, observations in the ATLAS $c$ (cyan) and $o$ (orange) bands are also included in the figure. Notably, the Mephisto observations capture a very early detection of SN~2026dyr in the $r$ band. During the early phase (within the first $\sim40$~days), the Mephisto data provide excellent coverage across all bands from $u$ to $z$, whereas the ATLAS detections are exclusively in the $o$ band, with no detections in the other bands (the $c$ bands are only available in the late phase where $t\gtrsim80$~days). This comprehensive coverage enables an accurate determination of the early color evolution and provides the multi-band data necessary to construct the spectral energy distribution (SED) as well as the bolometric light curve. We estimate the explosion epoch by averaging the epochs of the last non-detection and the first detection in the $r$ band, yielding an explosion time of MJD~$61092.4\pm0.5$. This result is more precise and earlier than the estimate derived from the ATLAS $c$ and $o$-band data, which yields MJD~$61093.5\pm3.0$.

Our dataset captures the plateau phase of SN~2026dyr; however, due to instrumental limitations, observations ceased at the end of this phase. Therefore, we utilized the analytical function provided by \cite{Valenti2016MNRAS.459.3939V} with slight modifications to fit our data. The fit yields a transitional time of $t_{\rm PT}=\mathrm{MJD}~61201.4\pm1.2$. Following the definitions of \cite{SN2016B_Dastidar2019MNRAS.486.2850D} and \cite{Anderson2014ApJ...786...67A}, we assume that the end-of-plateau epoch, $t_{\rm end}$, corresponds to the time when the apparent magnitude drops by $m_{\rm th}$ from the plateau, which we set to $0.1$~mag here. However, the result is not sensitive to this parameter, as the drop occurs on a relatively short timescale. Based on the applied model, the termination of the plateau can be expressed as:
\begin{equation}
    t_{\rm end} = t_{\rm PT} - w_0 \cdot \ln\left(\frac{a_0 - m_{\rm th}}{m_{\rm th}}\right),
\end{equation}
where $a_0$ is the amplitude of the magnitude drop after the plateau, and $w_0$ represents the duration of this drop. Applying our best-fit parameters yields $t_{\rm end} = \mathrm{MJD}~61182.2$. Consequently, the derived optically thick phase duration is 89.8~days, which is consistent with typical SNe~IIP \citep{Martinez2022A&A...660A..40M}.

Furthermore, we perform piecewise linear fits to the apparent light curves to derive the decline rates $s_1$, $s_2$, and $s_3$. Because observational coverage is limited after the plateau phase, we focus exclusively on the early decline rate ($s_1$) and the plateau decline rate ($s_2$). We find that $s_1$ in the $g$ band is $0.0062\pm0.0019\,\mathrm{mag\,(100d)^{-1}}$, while in the $r$ band it is $0.0091\pm0.0004\,\mathrm{mag\,(100d)^{-1}}$. These values are significantly flatter than those of typical Type IIP SNe exhibiting shock-cooling emission \citep[e.g.,][]{SN2018gj_Teja2023ApJ...954..155T, SN2017eaw_Szalai2019ApJ...876...19S}, although a brief shock-cooling phase remains evident. The $s_2$ values in the $g$ and $r$ bands are nearly zero. This behavior mirrors observations in other filters, including the ATLAS $o$ band, and aligns well with the standard evolution of SNe~IIP.

\subsection{Comparison of SN~2026dyr with other SNe~II events}

\begin{figure}
\centering
\includegraphics[width=0.9\linewidth]{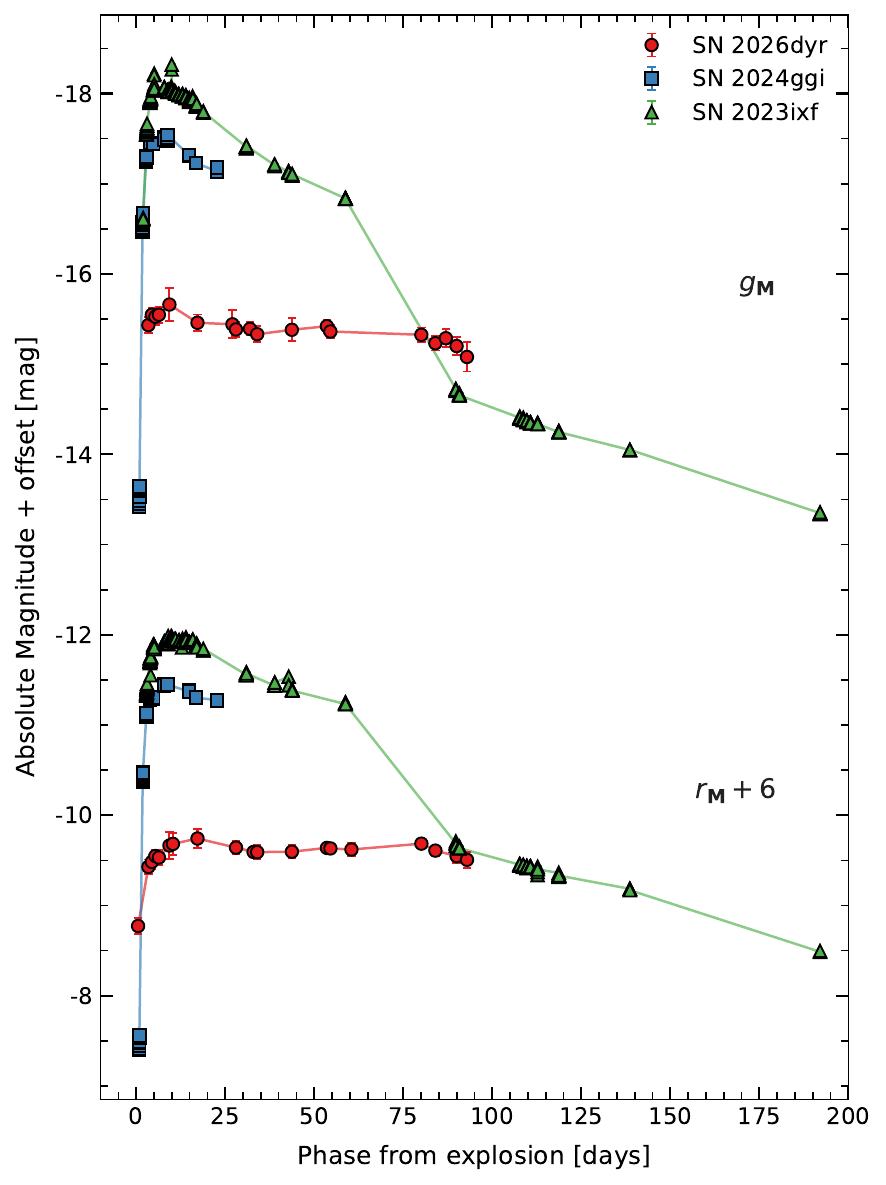}
\caption{Absolute magnitude evolution of SN~2026dyr compared with SN~2023ixf and SN~2024ggi in Mephisto $g$ and $r$ bands. The magnitudes are in the AB system. Different objects are displayed in different colors and symbols. A shift is applied in the $r$-band for clarity.}
\label{fig:26dyr:abslc}
\end{figure}

We compare the absolute light curves of SN~2026dyr with those of other Type II events (SNe 2023ixf, 2024ggi) well monitored by the Mephisto survey at very early epochs. The results are presented in Figure~\ref{fig:26dyr:abslc}. All light curves have been corrected for extinction and converted to absolute magnitudes using their respective distance moduli. We note that in the $g$ and $r$ bands, SN~2026dyr exhibits faint shock-cooling emission, characterized by a slight bump within the first $\sim30$~days. While similar profiles are observed in SN~2023ixf and SN~2024ggi, SN~2026dyr is approximately $2$~mag fainter during this early phase. The elevated early-time luminosity of SN~2023ixf and SN~2024ggi is primarily attributed to interaction with circumstellar material (CSM) \citep{Yang2024, Chen2024}, driven by strong pre-supernova winds from the red supergiant (RSG) progenitors. The comparative faintness of SN~2026dyr suggests that its immediate circumstellar environment is relatively clean, lacking prominent CSM signatures. Consequently, the pre-supernova activity for the progenitor of SN~2026dyr must have been significantly more quiescent than the activity seen in SN~2023ixf and SN~2024ggi, implying a pre-supernova mass-loss rate of $\dot M\lesssim10^{-5}\,\mathrm{M_{\odot}\,yr^{-1}}$ \citep[e.g.,][]{Sarin2025MNRAS.544.2653S}. However, we noticed that since the filter system of Mephisto is different from the SDSS $ugriz$ bands \citep[see Figure 1 in ][]{Yang2024}, it is not easy to directly perform comparisons to other well-studied events. But since the survey of Coma is established now, it enables us to enlarge the dataset of Mephisto for these events.

\begin{figure}
\centering
\includegraphics[width=0.9\linewidth]{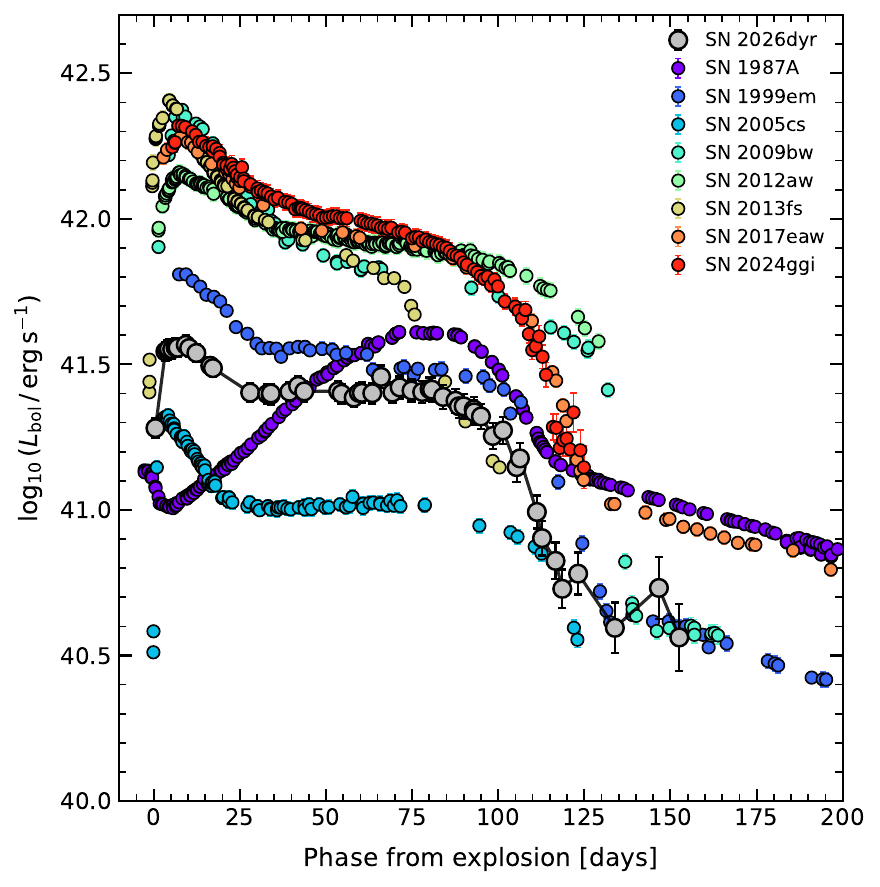}
\caption{Pseudo-bolometric light curve of SN~2026dyr is compared with other well-studied SNe~II events. The phases are calculated from the explosion epoch of the particular event. 
}
\label{fig:26dyr:bollc}
\end{figure}

We constructed the pseudo-bolometric light curve of SN~2026dyr by directly integrating the fluxes across the optical bands (from $u$ to $z$, corresponding to $L_{\rm UBVRI}$) using the \textsc{superbol} code \citep{Nicholl2018}. We slightly modified this code to incorporate the filter system of the Mephisto survey, following the method described by \cite{Chen2024}. We chose the $r$ band and the ATLAS $o$ band as reference filters for the interpolation. Because the color evolution of SNe~IIP proceeds very slowly at late epochs ($t\gtrsim100$~days) \citep[e.g.,][]{Martinez2022A&A...660A..40M}, the late-time fluxes were extrapolated under the assumption of constant color. Comparisons of the pseudo-bolometric light curves are presented in Figure~\ref{fig:26dyr:bollc}. SN~2026dyr exhibits early-phase shock-cooling emission, which is consistent with the behavior of typical SNe~IIP. However, compared to other SNe~IIP, the shock-cooling phase of SN~2026dyr is notably brief. The plateau luminosity of SN~2026dyr is $\log(L_{\rm bol}/\mathrm{erg\,s^{-1}})\simeq41.4$. This value is comparable to normal events like SN~1999em \citep{Elmhamdi2003MNRAS.338..939E}, but it is significantly fainter than luminous events such as SN~2017eaw and SN~2024ggi \citep{SN2017eaw_Szalai2019ApJ...876...19S, Ertini2025A&A...699A..60E}. Conversely, SN~2026dyr is notably brighter than low-luminosity events like SN~2005cs, which has a plateau luminosity of $\log(L_{\rm bol}/\mathrm{erg\,s^{-1}})\simeq41.0$. These features indicate that SN~2026dyr is likely a moderately energetic event accompanied by only marginal CSM interaction.

The plateau duration of SN~2026dyr is consistent with those of typical SNe~IIP. After hydrogen recombination concludes at $\sim80$~days, the luminosity drops abruptly from $\log(L_{\rm bol}/\mathrm{erg\,s^{-1}})\simeq41.4$ to $\log(L_{\rm bol}/\mathrm{erg\,s^{-1}})\simeq40.7$, entering the tail phase powered by radioactive decay. During this radioactive tail phase, the luminosity of SN~2026dyr is similar to the luminosities of SN~1999em and SN~2009bw, but notably lower than those of SN~1987A and SN~2017eaw. Because the late-time luminosity of Type II SNe can be approximated as the instantaneous deposition of radioactive decay energy \citep{Ravi2025ApJ...982...12R, Zhao2026arXiv260710671Z}, we estimate that the synthesized $^{56}$Ni mass for SN~2026dyr is comparable to the masses inferred for SN~1999em and SN~2009bw, which shared the same luminosity compared to the radioactive tail of SN~2026dyr. The detailed discussion of this feature can be found in Sect~\ref{LC-model}.

\subsection{Light curve modeling of SN~2026dyr}\label{LC-model}

We constructed the full bolometric light curve of SN~2026dyr by fitting blackbody functions to the spectral energy distributions using the \textsc{superbol} code \citep{Nicholl2018}. At mid and late times, the bolometric light curve of SN~2026dyr can be modeled simply by combining contributions from radioactive decay and recombination processes \citep[see][]{ArnettFu1989}. At early times, we observed a bump in the light curve, which likely originates from shock cooling emission and/or interaction with a thin CSM. Therefore, we utilized the two-component semi-analytical model presented by \citet{NagyVinko2016}, which combines a dense, massive core component with an extended thin shell. Radioactive decay only occurs within the core component, whereas the outer shell contributes to the luminosity primarily through the diffusion of initial thermal energy. Given that the environment is hydrogen-rich, we set the opacity to $\kappa=0.4\,\mathrm{cm^{2}\,g^{-1}}$ and the recombination temperature to $T_{\rm rec}=6000\,\mathrm{K}$, values that are typical for such environments. Assuming that opacity is dominated by electron scattering from ionized hydrogen, regions where $T<T_{\rm rec}$ can be considered completely optically thin ($\kappa=0$). This assumption allows $T_{\rm rec}$ to trace the evolution of the photosphere. We assume a uniform density profile for the core, whereas for the shell, we adopt a density profile of $\rho\propto r^{-2}$, consistent with a steady stellar wind or a low-density outer shell \citep{Nagy2014, NagyVinko2016}. 

\begin{figure}
\centering
\includegraphics[width=0.9\linewidth]{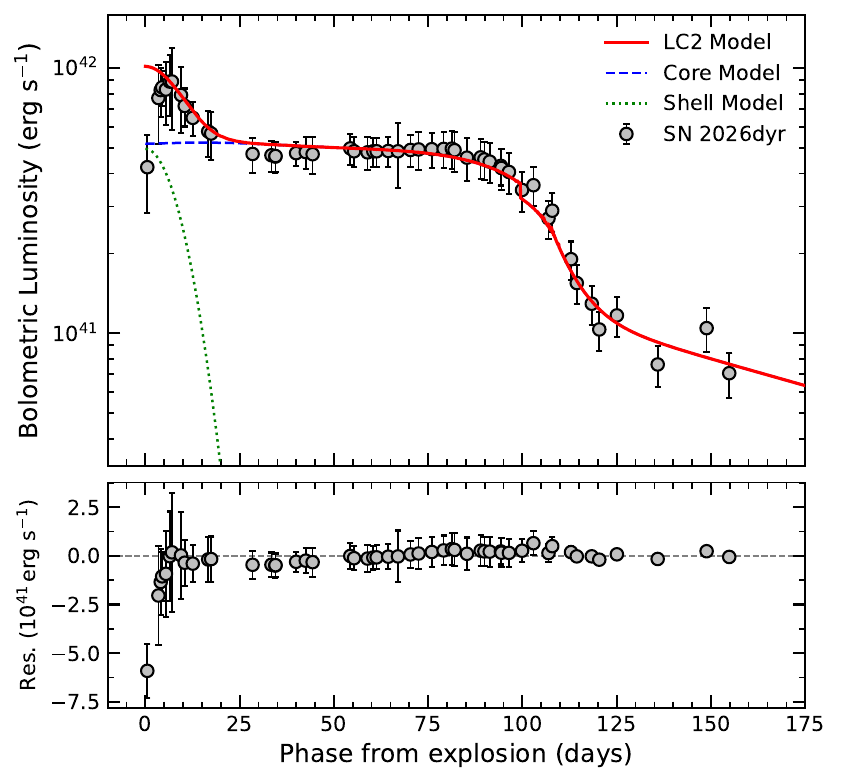}
\caption{Two-component light curve model of SN~2026dyr (see Section~\ref{LC-model}). \textit {Upper panel}: light curve models overplotted with the observational data of SN~2026dyr. The combination of the two components is plotted in a red solid line, and the other components are plotted with blue (dashed) and green (dotted) colors. \textit {Lower panel}: residuals of the fitting in the two-component fitting.}
\label{fig:26dyr:lcmdl}
\end{figure}

Because the parameter space of the two-component model is high-dimensional and the degeneracies between parameters are strong \citep[e.g., $E_{\rm th}$ and $R_{0}$; $E_{\rm k}$ and $M_{\rm ej}$; see][]{Jager2020MNRAS.496.3725J}, we adjusted the parameters within a reasonable range and found the best-match model following the method described by \citet{Nagy2014}. The best-fit results are presented in Figure~\ref{fig:26dyr:lcmdl}. This model yields an ejecta mass of $M_{\rm ej}=9.0\,M_{\odot}$ and a kinetic energy of $E_{\rm k}=1.4\,\mathrm{foe}$. The initial thermal energy is $E_{\rm th}=0.5\,\mathrm{foe}$, which is consistent with the majority of the SNe~IIP sample presented by \citet{NagyVinko2016}. The radius of the core in this scenario is $R_{0}\simeq 245\,R_{\odot}$. This value indicates that the progenitor of SN~2026dyr is relatively compact, similar to SN~2005cs \citep[$\sim 172\,R_\odot$;][]{NagyVinko2016}, and suggests that the progenitor could be a yellow supergiant. Within this framework, the ejecta velocity is $v_{\rm ej}=(E_{\rm k}/0.3M_{\rm ej})^{1/2} \simeq 5000\,\mathrm{km\,s^{-1}}$. This velocity is consistent with typical SNe~IIP, although it falls near the less energetic end of the distribution. The nickel mass derived from this fit is $M_{\rm Ni}=0.02\,M_{\odot}$, which aligns with our previous discussion. Furthermore, the model suggests a low-density shell component with a radius of $R_{\rm env}=287\,R_{\odot}$ and a mass of approximately $M_{\rm sh}=0.250\,M_{\odot}$. These findings point to a low-mass and compact shell, matching the observational data and resembling the profile of the less energetic, low-luminosity event SN~2005cs.

The only difference arises in the early-phase data, where the residual reaches approximately $-6\times10^{41}\,\mathrm{erg\,s^{-1}}$. A similar early-time discrepancy was also noted during the modeling process by \citet{Chen2024}, suggesting that an additional mechanism must be implemented to describe the early-phase evolution accurately. Furthermore, the model provided by \citet{NagyVinko2016} cannot explain the very early evolution of the light curve, particularly the shock breakout and shock cooling phases. Most SNe~IIP progenitors are surrounded by a thin CSM driven by strong red supergiant (RSG) winds during late stellar evolutionary stages \citep[see][]{Yang2024, Sarin2025MNRAS.544.2653S}; interaction with this material could explain the observed light curve evolution at early times. However, our target is sub-luminous and lacks prominent signatures of CSM interaction, such as narrow emission lines at early times. Moreover, the host environment of SN~2026dyr is extremely metal-poor, with $\log(Z/Z_\odot)=-0.813$\footnote{See \url{https://blast.scimma.org/transients/2026dyr/}}. Progenitors in such environments typically exhibit weak stellar winds, making it difficult to form the dense CSM required by semi-analytical CSM interaction models.

As a consistency check, we utilize a semi-quantitative method to calculate the $^{56}$Ni mass by comparing the radioactive tail of SN~2026dyr to that of SN~1987A. The late-time bolometric luminosity powered by radioactive decay can be described by the following function \citep[e.g.,][]{2012Chatzopoulos, Ravi2025ApJ...982...12R}:
\begin{equation}
    L_{\rm bol} = M_{\rm Ni} \left[ (\epsilon_{\rm Ni} - \epsilon_{\rm Co})e^{-t/\tau_{\rm Ni}} + \epsilon_{\rm Co}e^{-t/\tau_{\rm Co}} \right] \;.
\end{equation}
For $t \gg \tau_{\rm Ni}$, as is the case for the tail of SN~2026dyr, the contribution from $^{56}$Ni decay is negligible, and the luminosity follows $L_{\rm bol} \propto e^{-t/\tau_{\rm Co}}$. In logarithmic space, this reduces to a simple linear function, $\log L_{\rm bol} \propto -t$. Because the late-phase data of SN~2026dyr have limited temporal coverage and noticeable scatter, we cannot fully apply the method described by \cite{Ravi2025ApJ...982...12R}, as the $\gamma$-ray leakage effect is difficult to constrain. Therefore, we assume full trapping and simply fix the decline rate of the radioactive tail to match the expected $^{56}$Co decay rate. 
Thus, the $^{56}\mathrm{Ni}$ mass can be estimated by comparing the luminosity of SN~2026dyr to that of well-studied events such as SN~1987A. Given that the late-time data coverage is sparse and noisy, we instead utilize the intercept of the radioactive decay tail in logarithmic luminosity space for both objects, which provides a more robust estimate in this case.
Adopting $M_{\rm Ni,\,87A}=0.075\,\mathrm{M_\odot}$ \citep[e.g., see][]{Spiro2014MNRAS.439.2873S}, our fit yields $M_{\rm Ni}=0.028\pm0.005\,\mathrm{M_\odot}$ for SN~2026dyr, which is highly consistent with our previous qualitative estimation. Together, these features suggest that SN~2026dyr represents a relatively weak, low-energy explosion that synthesized a modest amount of radioactive material and consequently produced a low-luminosity light curve.

\section{Expected Transient Rates and Pilot Coma Survey Yield}
\label{sec:event_rate}

\begin{table*}
\small
\centering
\caption{Stellar Mass and Star Formation Rate in the Coma Field}
\label{tab:performance}

\resizebox{\textwidth}{!}{%
\begin{tabular}{lcccccccccc}
\toprule
\multirow{2}{*}{Maximum Distance} 
& \multicolumn{5}{c}{\textbf{Stellar Mass}} 
& \multicolumn{5}{c}{\textbf{Star Formation Rate}\hspace{1em}} \\
\cmidrule(lr){2-6} \cmidrule(lr){7-11}
& \textbf{$M_{\rm C}$} 
& \textbf{$M_{\rm bk}$} 
& \textbf{$M_{\rm med}$} 
& \textbf{$M_{\rm C}/M_{\rm bk}$} 
& $M_{\rm C}/M_{\rm med}$ 
& \textbf{$\rm SFR_{\rm C}$} 
& \textbf{$\rm SFR_{\rm bk}$} 
& \textbf{$\rm SFR_{\rm med}$} 
& \textbf{$\rm SFR_{\rm C}/SFR_{\rm bk}$} 
& $\rm SFR_{\rm C}/SFR_{\rm med}$ \\
\midrule

$\leq$ 100 Mpc 
& $1.11\times 10^{13}$ 
& $4.63\times 10^{11}$ 
& $8.77\times 10^{10}$ 
& 23.97 
& 126.57 
& 138.06 
& 9.93 
& 3.78 
& 13.90 
& 36.52 \\

$\leq$ 150 Mpc 
& $2.77\times 10^{13}$ 
& $1.77\times 10^{12}$ 
& $8.21\times 10^{11}$ 
& 15.65 
& 33.74 
& 285.82 
& 34.16 
& 21.90 
& 8.37 
& 13.05 \\

$\leq$ 200 Mpc 
& $3.04\times 10^{13}$ 
& $3.86\times 10^{12}$ 
& $2.34\times 10^{12}$ 
& 7.88 
& 12.99 
& 366.84 
& 85.64 
& 66.39 
& 4.28 
& 5.53 \\

\bottomrule
\end{tabular}%
}

\vspace{0.5em}
\parbox{\textwidth}{%
\scriptsize
\textbf{Note:} All quantities are linear. Column (1) gives the cumulative distance limit. $M_{\rm C}$ (in $\rm M_{\odot}$) and $\rm SFR_{\rm C}$ (in $\rm M_{\odot}\,yr^{-1}$) are the total stellar mass and total star formation rate of galaxies located within the Coma cluster region, which is defined as the union of the ten sub-fields of the Coma field file (rectangular RA/Dec ranges). $M_{\rm bk}$ and $M_{\rm med}$ (in $\rm M_{\odot}$) are the expected mean and median background stellar masses within the Coma region, obtained by multiplying the corresponding background surface densities by the fixed Coma area $A_{\rm Coma}=20\,\rm deg^{2}$. The mean background surface density is the total stellar mass of galaxies outside the Coma region divided by the full-sky area $A_{\rm sky}=41253\,\rm deg^{2}$, and the median background surface density is measured from the $1\fdg83$ pixelized full-sky map with empty pixels counted as zero and divided by the pixel area of $3.357\,\rm deg^{2}$. $\rm SFR_{bk}$ and $\rm SFR_{med}$ (in $\rm M_{\odot}\,yr^{-1}$) are the corresponding expected star-formation-rate values within the Coma region. The ratio columns are dimensionless enhancement factors of the Coma region
relative to the background expectation.
}
\end{table*} 

\begin{figure*} 
    \centering
    \includegraphics[width=0.3\linewidth]{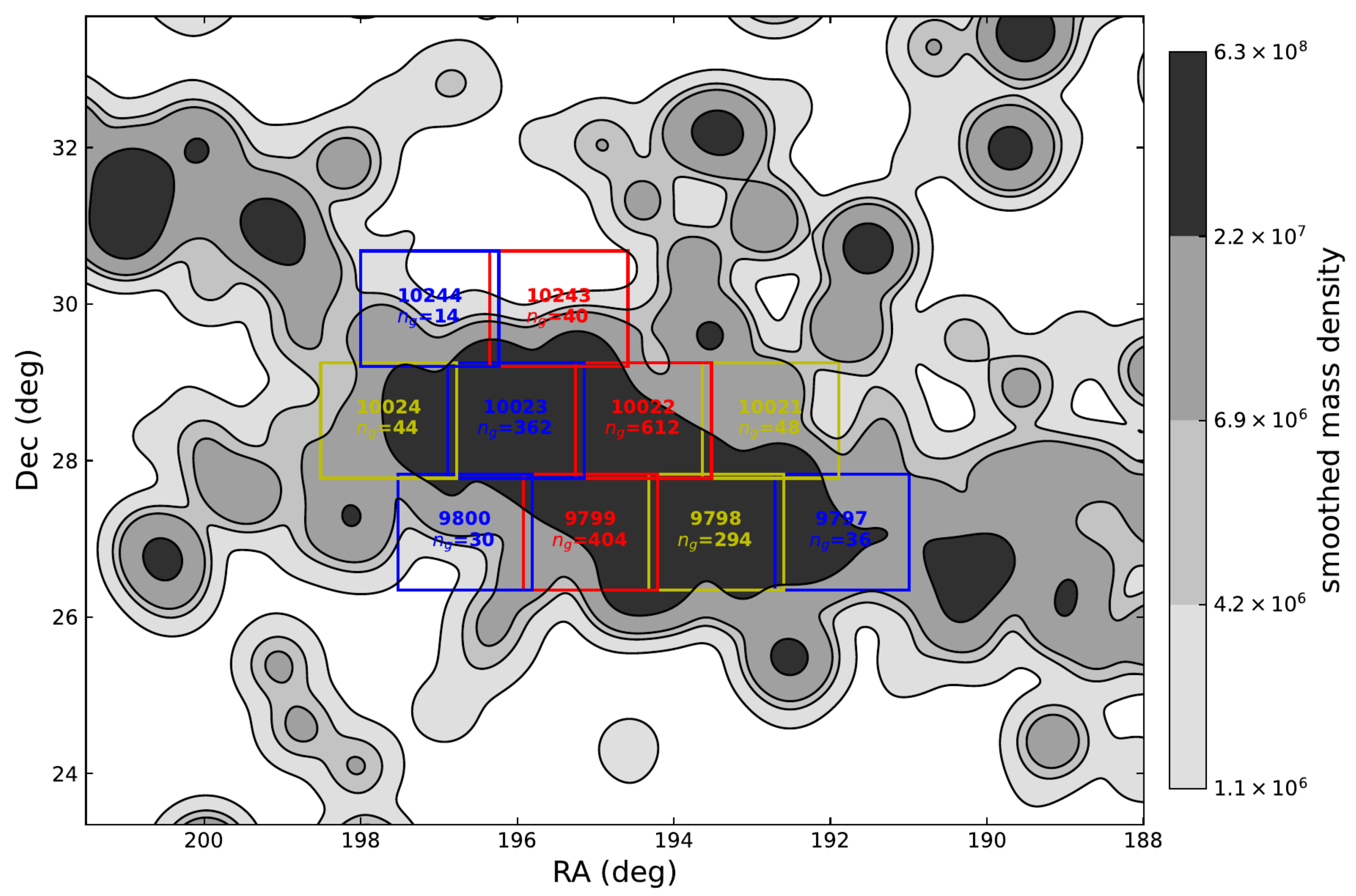}
    \includegraphics[width=0.3\linewidth]{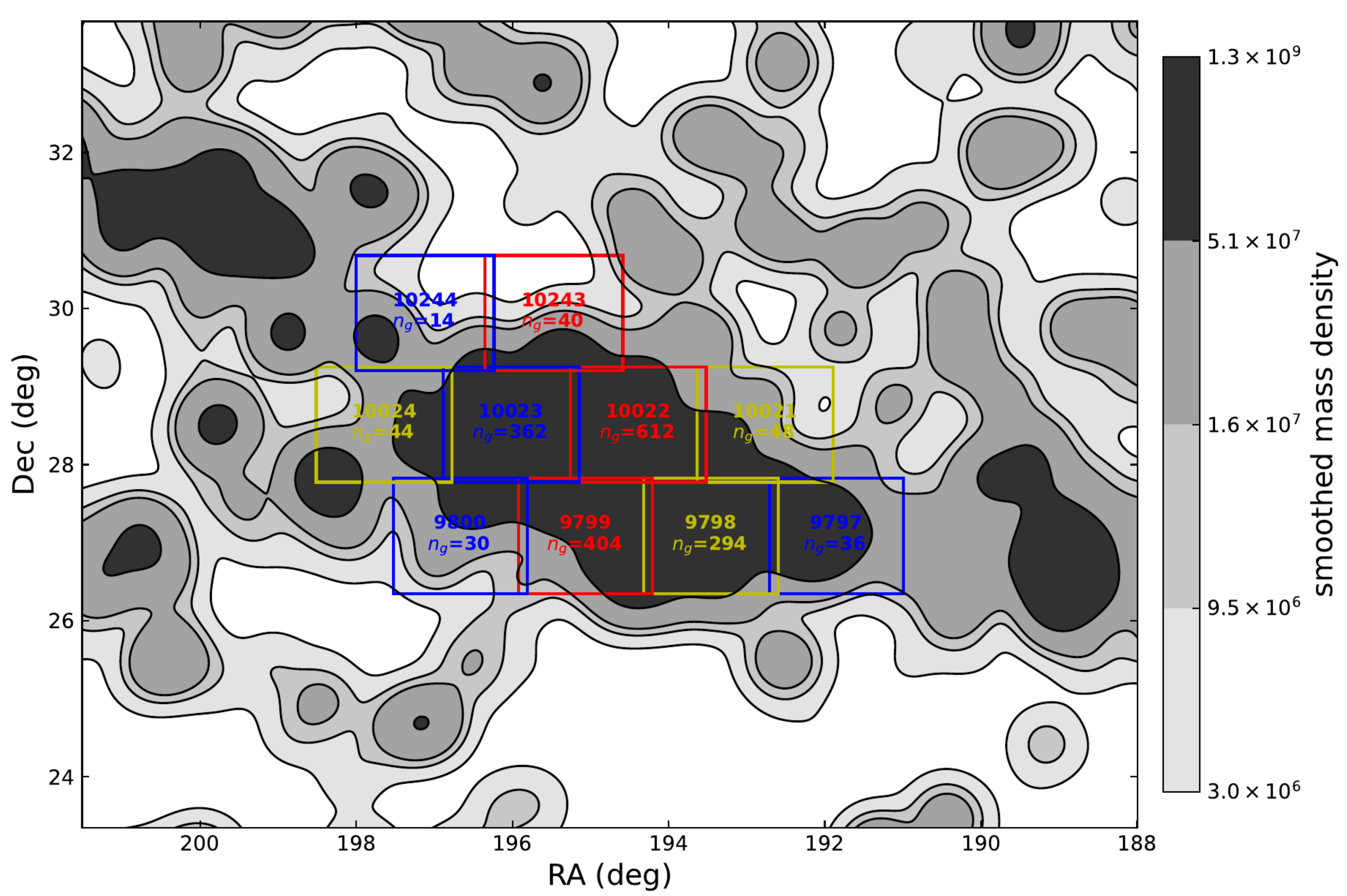}
    \includegraphics[width=0.3\linewidth]{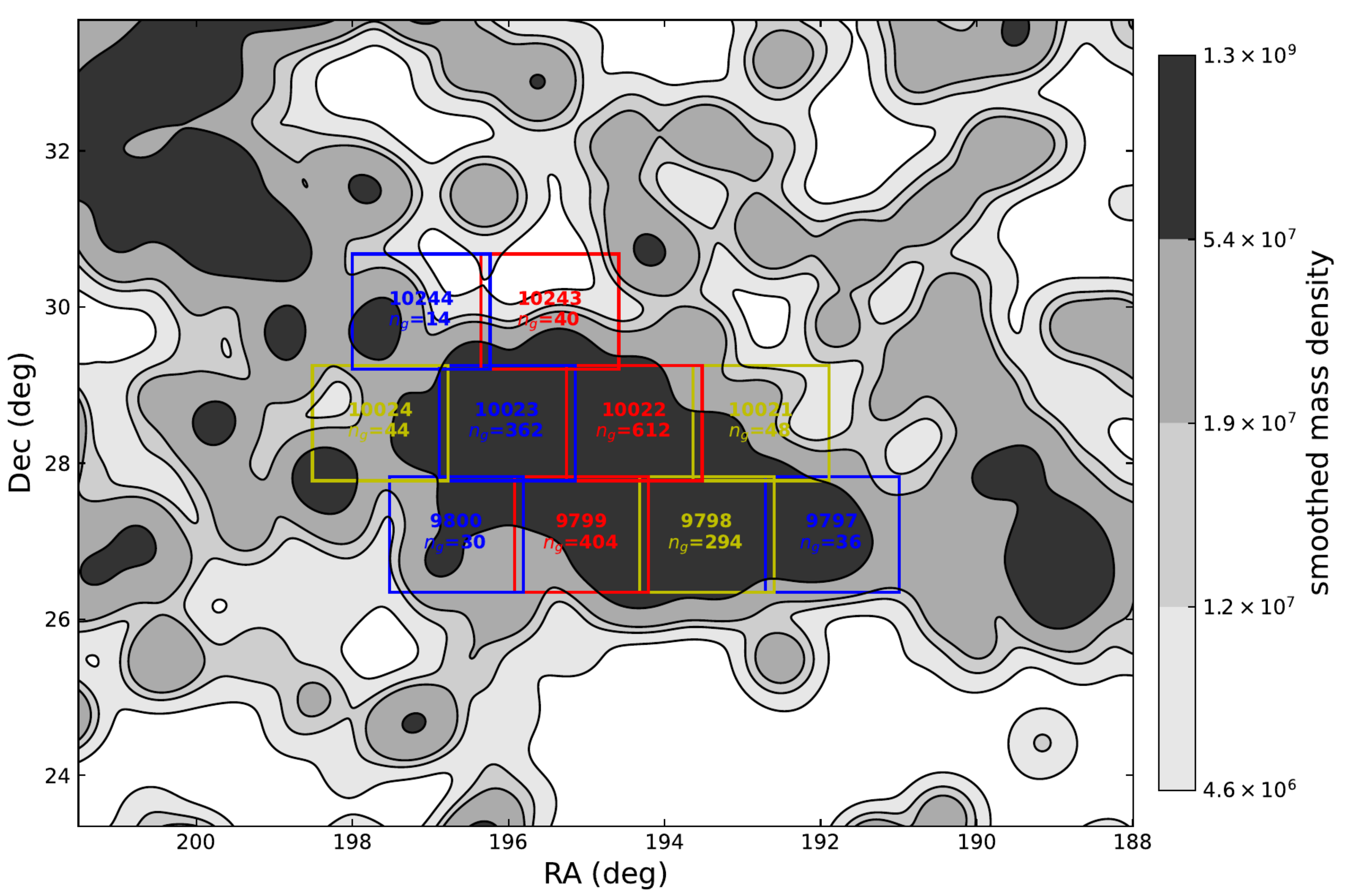}
    \includegraphics[width=0.3\linewidth]{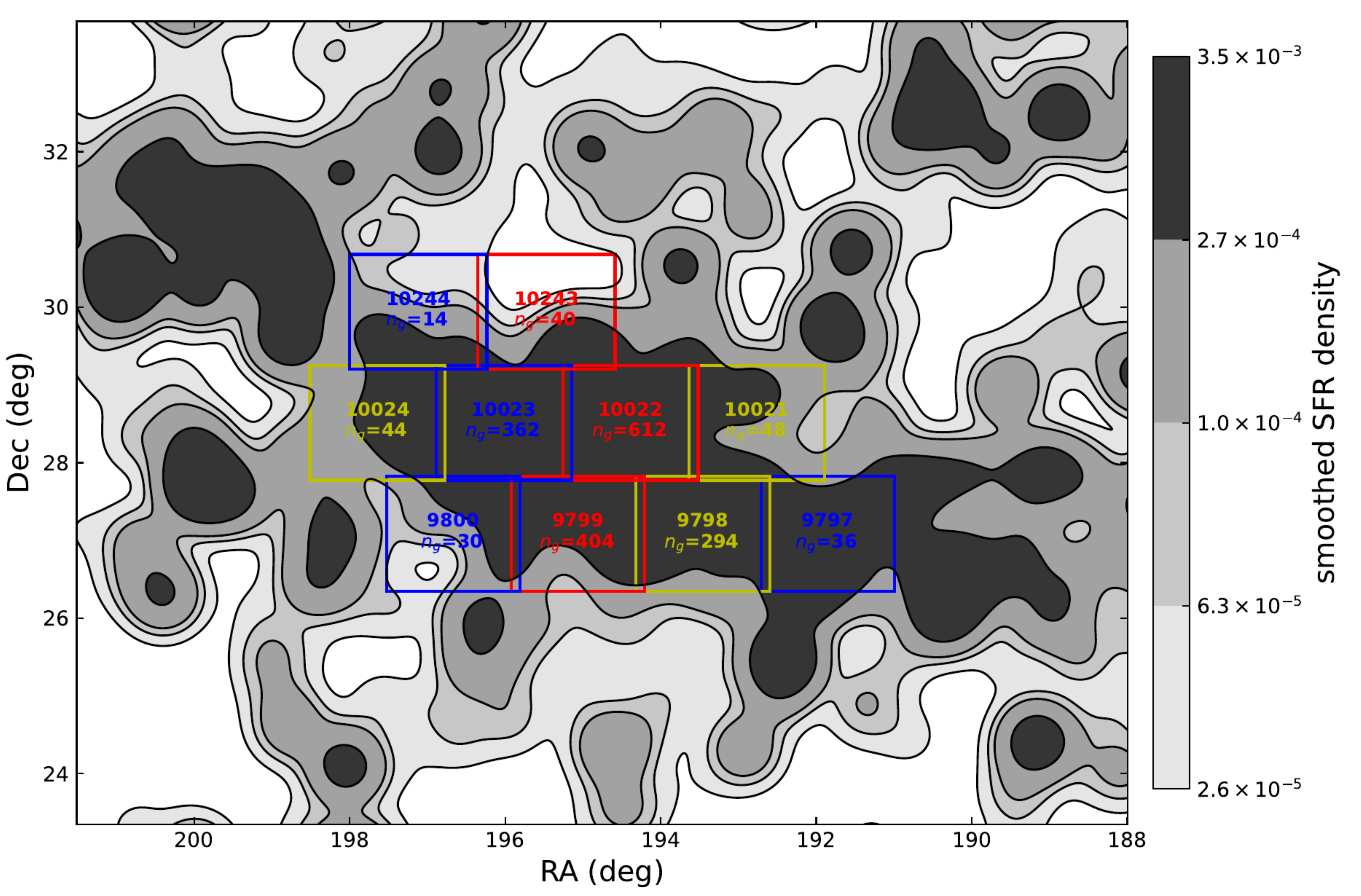}
    \includegraphics[width=0.3\linewidth]{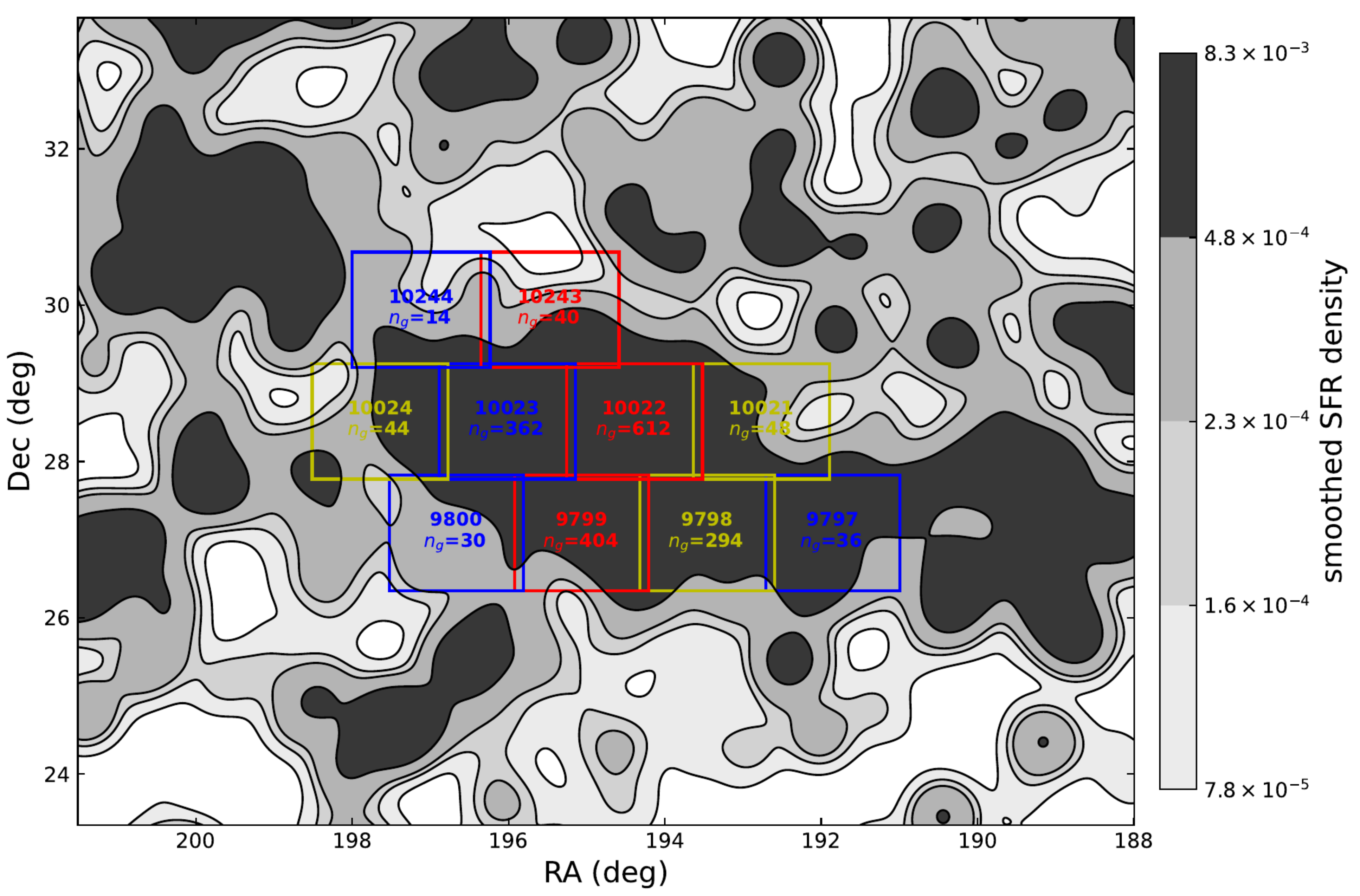}
    \includegraphics[width=0.3\linewidth]{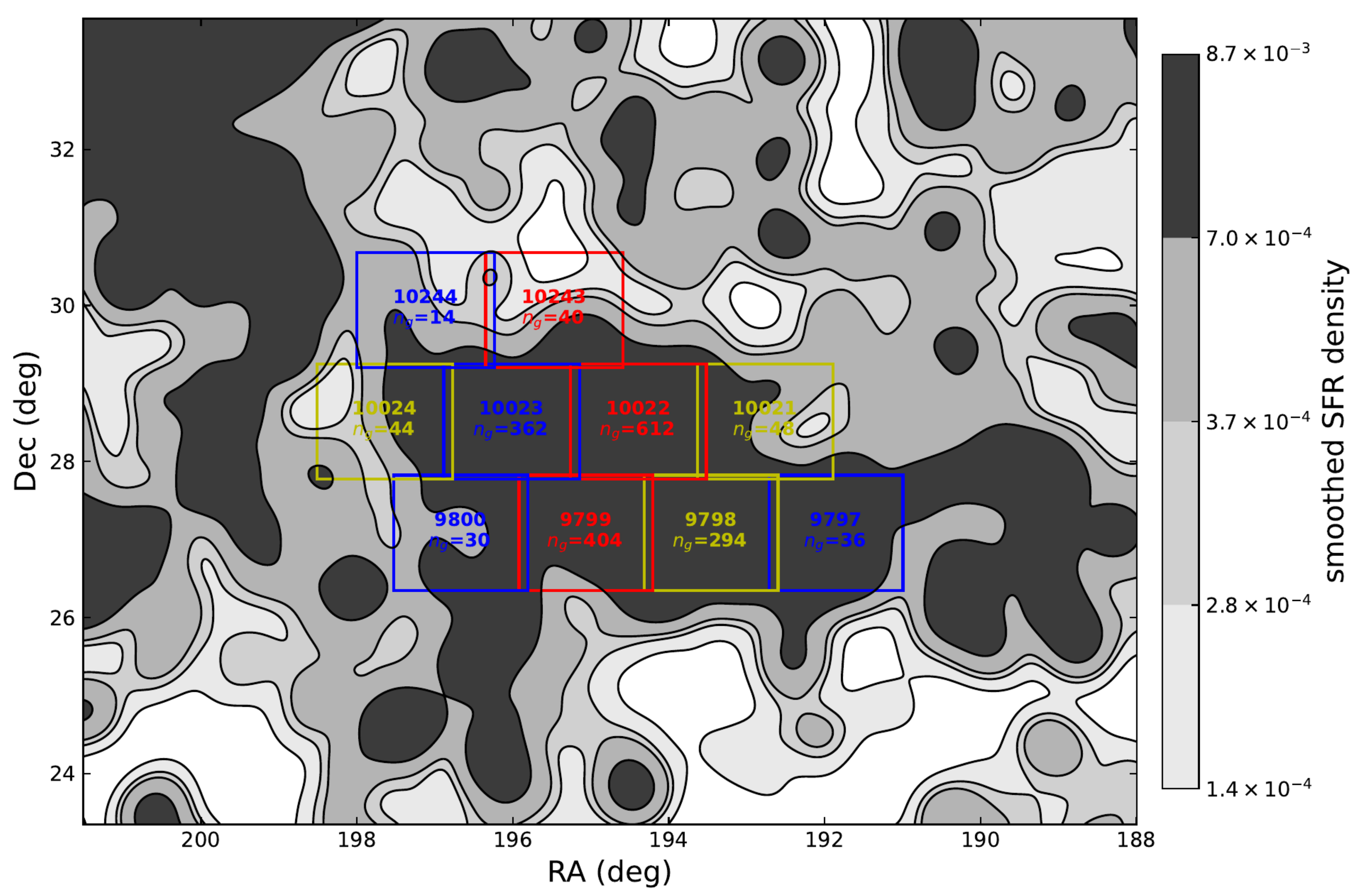}
    \caption{Stellar Mass (top panels) and Star Formation Rate (bottom panels) in the Coma field at different distance thresholds of 100 Mpc (left panels), 150 Mpc (middle panels), and 200 Mpc (right panels). The blue, yellow, and red color regions in the plot indicate different coverage areas during the sky survey by Mephisto.}\label{fieldrate}
\end{figure*}

To quantitatively justify our strategy of targeting the Coma cluster and to evaluate Mephisto's true detection efficiency, this section compares theoretical expectations with our observational reality. We map the local overdensity of stellar mass and star formation rate to predict event rates for various transient classes, and benchmark these forecasts against our pilot survey yield.
Unlike the highly isotropic distribution of transients at cosmological distances, the local universe exhibits pronounced large-scale structures, leading to a significant spatial anisotropy in transient event rates. To optimize our survey efficiency and quantitatively evaluate our results, we leverage the recent comprehensive local universe sky maps generated by \citet{Cheng26}. By mapping the distributions of stellar mass and star formation rate (SFR) across various distance thresholds, their work provides a robust framework for characterizing the highly anisotropic clustering of nearby extragalactic transients. 

Following the procedure of \cite{Cheng26}, we generate data for the sky distributions of stellar mass and SFR with a pixel size of $1.72'$. For a given angular resolution, the celestial sphere can be divided into equal-area curvilinear quadrilaterals, where the pixel size is equal to the angular resolution of the grid \citep{HEALPix2005}. In Figure \ref{fieldrate}, based on the newly produced data, we calculate the smoothed projected stellar mass / SFR density fields toward the Coma cluster for luminosity distance thresholds of $D_{\rm L}$ = 100, 150, and 200 Mpc, together with the ten subfields. The plotting window is the union of the ten subfields enlarged by $3^{\circ}$ (RA: $188^{\circ} - 201.5^{\circ}$, Dec: $23.3^{\circ} - 33.7^{\circ}$). 
The maps are produced as follows. 
(1) Valid pixels are selected within a $2^{\circ}$ buffer around the window, excluding empty pixels.
(2) Using the linear mass (or linear SFR), as weights, we build a weighted histogram on a $0.01^{\circ}$ grid and smooth it with a Gaussian kernel of width $\sigma=0.3^{\circ}$, yielding a continuous smoothed projected density field $\rho(x)$ at sky position $x$, which is equivalent to a Gaussian kernel-density estimate. 
(3) Contour levels are defined by the enclosed weight fraction: the threshold $\tau_f$ satisfies $\int_{\rho \geqslant \tau_f} \rho dA / \int \rho dA=f $, where $dA$ is the sky area element, with $f = {99\%, 95\%, 90\%, 68\%}$; from outside inward, the four contours therefore enclose 99\%, 95\%, 90\%, and 68\% of the total mass (or SFR) within the window, analogous to credible-level contours of a posterior distribution; the upper edge of the innermost filled contour band is set to 1.05 times the peak of $\rho$ so that the peak region is fully enclosed (and coloured) within the 68\% contour band.

Based on the detailed mapping discussed above (as summarized in Table~\ref{tab:performance}), we extract the integrated astrophysical parameters within our 20-square-degree Coma cluster survey footprint. Integrating out to a luminosity distance of $D_{\rm L} \leq 150$\,Mpc, thus encompassing both the dense Coma cluster core and the rich foreground/background filaments along the line of sight, the total star formation rate is strictly determined to be $\text{SFR}_{150} = 285.82 \, \rm M_\odot\,\text{yr}^{-1}$, with an integrated stellar mass of $M_{*, 150} = 2.77 \times 10^{13} \, \rm M_\odot$. Notably, this specific volume presents a staggering structural overdensity: the stellar mass and SFR within the Coma field are enhanced by factors of $\sim 15.7$ and $\sim 8.4$ relative to the mean background expectation, respectively.

We can systematically estimate the theoretical event rates of CCSNe and Type Ia supernovae (SNe Ia) in this highly overdense volume by scaling them against the baseline parameters of the Milky Way (MW). Since CCSNe are the direct endpoints of massive, short-lived stars, their volumetric rate strictly traces recent star formation activity. Adopting the MW star formation rate of $\text{SFR}_{\text{MW}} \simeq (1.65 \pm 0.19) - (2.0 \pm 0.7) \, \rm M_\odot\,\text{yr}^{-1}$ \citep{Licquia15,Elia22} and a corresponding galactic CCSN rate of $\mathcal{R}_{\text{MW}}^{\text{CC}} \simeq (1.63 \pm 0.46) - (1.9 \pm 1.1)\,\text{century}^{-1}$ \citep{Diehl06,Rozwadowska21}, the expected CCSN rate in our target field is given by:
\begin{equation}
    \mathcal{R}_{150}^{\text{CC}} = \mathcal{R}_{\text{MW}}^{\text{CC}} \left( \frac{\text{SFR}_{150}}{\text{SFR}_{\text{MW}}} \right) \simeq 1.2 - 5.9 \, \text{yr}^{-1}.
\end{equation}
Conversely, the SN Ia rate is largely proportional to the total stellar mass, originating from older stellar populations. Utilizing the MW stellar mass of $M_{*, \text{MW}} \simeq (5.43 \pm 0.57) - (6.08 \pm 1.14) \times 10^{10} \, \rm M_\odot$ \citep{Licquia15,McMillan17} and a Galactic SN Ia rate of $\mathcal{R}_{\text{MW}}^{\text{Ia}} \simeq (0.28 \pm 0.06) - (0.80\pm 0.23)\,\text{century}^{-1}$ \citep{Li11,Munday25,2025A&A...698A.306M}, the expected SN Ia rate in the targeted volume is:
\begin{equation}
    \mathcal{R}_{150}^{\text{Ia}} = \mathcal{R}_{\text{MW}}^{\text{Ia}} \left( \frac{M_{*, 150}}{M_{*, \text{MW}}} \right) \simeq 0.84 - 5.9 \, \text{yr}^{-1}.
\end{equation}

We compare these theoretical forecasts with the empirical yield from our pilot operations. During the initial six-month observational window, the Mephisto Coma field survey discovered a total of 9 confirmed supernovae. After excluding two events that were spectroscopically confirmed to be background transients beyond the 150 Mpc threshold, we obtained a pure local sample of seven supernovae over a six-month period, one of which, SN 2026exy, lacks a redshift measurement. This sample is composed of 5 CCSNe and 2 SNe Ia. Assuming a constant discovery rate, this translates to an observed annual rate of $\sim 10\,\text{yr}^{-1}$ for CCSNe and $\sim 4\,\text{yr}^{-1}$ for SNe Ia in this localized field.

The empirical results exhibit a well structural agreement with the predictions. The fact that the observed CCSN rate surpasses the SN Ia rate perfectly reflects the integration of the $150$\,Mpc volume: while the core of the Coma cluster is dominated by passive early-type galaxies (producing SNe Ia), the extensive line of sight captures highly active, star-forming filaments, heavily boosting the SFR and, consequently, the CCSN yield. Furthermore, the observed rates ($\sim 10$ CCSNe and $\sim 4$ SNe Ia per year) are consistently higher than or close to the upper bounds of the theoretical predictions ($\sim 5.9$ and $\sim 5.9$ per year, respectively). This excess not only validates the extreme local overdensity mapped by \citet{Cheng26} but also definitively highlights the exceptional efficiency and completeness of Mephisto's high-cadence, multi-channel strategy in identifying transients.

Encouraged by the rich harvest of standard supernovae, we extend this physical framework to predict the occurrence rates of a broader spectrum of astrophysical transients within our target field. Distinct classes of transients exhibit varying dependencies on progenitor environments. Specifically, transients originating from massive stellar deaths (e.g., various CCSN subtypes, Long Gamma-Ray Bursts, and Pair-Instability Supernovae) scale proportionally with the SFR. In contrast, transients arising from compact object remnants (e.g., Binary Black Hole or Neutron Star mergers) scale primarily with the integrated stellar mass.

It is crucial to clarify the physical motivation for establishing two independent baselines ($\mathcal{R}_{150}^{\text{CC}}$ and $\mathcal{R}_{150}^{\text{Ia}}$) rather than simply scaling expected rates from a universal cosmic volumetric ratio. Cosmic volumetric rate densities are averaged over large cosmological volumes, effectively assuming a mean cosmic specific star formation rate (sSFR) or a mean cosmic stellar mass density. However, the Coma cluster represents a highly biased, dense environment dominated by massive, passive early-type galaxies. 
Consequently, applying a universal volumetric ratio to such an extreme overdensity would systematically bias the predictions. To achieve accurate forecasts, it is imperative to decouple these populations based on their progenitor physics: we strictly anchor transients with short delay times (driven by massive star deaths) to the local SFR, and transients with long delay times (arising from older compact objects) to the integrated stellar mass.

By adopting the local universe volumetric rate densities ($\rho$) derived by \citet{Briel22}, we extrapolate the expected event rates for any specific transient class $X$ in our Coma field ($\mathcal{R}_{150}^{X}$) according to the volumetric ratio relative to their respective physical proxy:
\begin{equation}
    \mathcal{R}_{150}^{X} = \mathcal{R}_{150}^{\text{Baseline}} \left( \frac{\rho_{X}}{\rho_{\text{Baseline}}} \right).
\end{equation}

\begin{table*}
\centering
\caption{Expected Transient Event Rates in the Mephisto Coma Field ($D_{\rm L} \leq 150$\,Mpc, 20 sq. deg.). Volumetric rate densities ($\rho$) are adopted from \citet{Briel22}. Expected event rates are derived by scaling the volumetric rates to the theoretical baselines calculated for CCSNe and SNe Ia based on the latest physical parameters from Table~\ref{tab:performance}.}
\label{tab:transient_rates}
\renewcommand{\arraystretch}{1.3}
\begin{tabular}{llcc}
\toprule
\textbf{Transient Class} & \textbf{Progenitor Proxy} & $\mathbf{\log_{10}(\rho)}$ & \textbf{Expected Event Rate} \\
 & & \textbf{(Gpc$^{-3}$ yr$^{-1}$)} & $\mathbf{\mathcal{R}_{150}}$ \textbf{(yr$^{-1}$)} \\
\midrule
\multicolumn{4}{c}{\textit{Transients Tracing Star Formation Rate}} \\
\midrule
Core-Collapse SNe (All) & SFR  & $4.92$ & $1.2 - 5.9$ \\
Type II SNe (non-IIP)   & SFR            & $4.20$ & $0.23 - 1.1$ \\
Type IIP SNe            & SFR            & $4.64$ & $0.63 - 3.1$ \\
Type Ib SNe             & SFR            & $4.15$ & $0.20 - 1.0$ \\
Type Ic SNe             & SFR            & $3.98$ & $0.14 - 0.68$ \\
Long Gamma-Ray Bursts   & SFR            & $2.29$ & $(2.8 - 13.8) \times 10^{-3}$ \\
Pair-Instability SNe    & SFR            & $0.82$ & $(1.0 - 4.7) \times 10^{-4}$ \\
\midrule
\multicolumn{4}{c}{\textit{Transients Tracing Stellar Mass}} \\
\midrule
Type Ia SNe             & Mass  & $4.25$ & $0.84 - 5.9$ \\
Binary Neutron Star (BNS)      & Mass            & $2.35$ & $(1.1 - 7.4) \times 10^{-2}$ \\
Binary BH-NS (BHNS)     & Mass            & $2.38$ & $(1.1 - 8.0) \times 10^{-2}$ \\
Binary Black Hole (BBH) & Mass            & $2.01$ & $(0.48 - 3.4) \times 10^{-2}$ \\
\bottomrule
\end{tabular}
\end{table*}

The comprehensive theoretical yields for the 20-square-degree Coma field ($D_{\rm L} \leq 150$\,Mpc) are summarized in Table~\ref{tab:transient_rates}. Unsurprisingly, standard core-collapse subtypes (e.g., SNe IIP, SNe Ib/c) are expected to occur at a frequency of several events per decade. More critically, the dense integration of stellar mass along this specific line of sight yields non-negligible rates for gravitational wave sources. For instance, we expect approximately $1.1 - 7.4$ binary neutron star (BNS) mergers and $1.1 - 8.0$ black hole-neutron star (BHNS) mergers per century within this precise footprint. 
With its 10-minute cadence and real-time multi-band color information, Mephisto is well suited for the rapid identification and characterization of the fast-evolving optical counterparts of BNS and BHNS mergers. In the event of a gravitational-wave trigger localized to this overdense cosmic structure, the high-cadence observations would provide an effective means of catching the early, blue emission from the associated kilonova and tracing its subsequent color evolution.

\section{Other Extreme Transients in the Coma Cluster}
\label{sec:broad_science}

While early supernova physics constitutes a primary driver for this survey, the 20-square-degree continuous monitoring of the Coma cluster unlocks several other high-impact astrophysical frontiers. The combination of dense galactic environments, rapid cadence, and real-time colors presents unique opportunities for discovering rare, fast-evolving phenomena.

\subsection{Tidal Disruption Events and AGN Variability}
The Coma cluster harbors thousands of massive galaxies, each potentially hosting a central supermassive black hole (SMBH). When a star wanders too close to an SMBH, it is violently torn apart, producing a Tidal Disruption Event (TDE). Early-stage TDEs are characterized by extremely blue optical continuum emission and peak on timescales of weeks. Mephisto's simultaneous $u-g-i$ coverage is exquisitely sensitive to the hot thermal emission of nascent TDEs. The instantaneous colors will allow us to immediately distinguish TDE candidates from standard supernovae or variable Active Galactic Nuclei (AGN) in real time. Furthermore, for the existing AGN population within the cluster, the strictly simultaneous multi-color light curves eradicate the intrinsic color noise introduced by sequential filter wheels, enabling high-precision optical reverberation mapping and the study of accretion disk thermal fluctuations.

Among the transients detected by Mephisto, we identified a possible candidate for a TDE or an AGN, registered on the TNS as AT~2025xfp \citep{Perley2025TNSTR3656....1P}. The coordinates of this detection are $\alpha = 13^{\rm h}05^{\rm m}53\fs108$, $\delta = +27\degr33\arcmin07\farcs49$ (J2000), which correspond to a spatial offset of approximately $0\farcs31$ from the center of the galaxy NGC~4960. This observation occurred $\sim100$~days after the initial discovery by the ZTF. However, because we only acquired a single frame for this target, we cannot definitively confirm whether the physical nature of the event is a genuine TDE or an AGN. Concurrent photometric data from ZTF near the epoch of our observation indicate a brightness of $\sim18.5$~mag in the $r$-band.

\subsection{Intra-cluster Transients and Ultra-Diffuse Galaxies}
A unique advantage of targeting a massive cluster like Coma is the substantial presence of Intra-Cluster Light (ICL)—diffuse stellar populations bound to the cluster potential rather than individual galaxies. Traditional transient surveys often miss or misclassify ``hostless'' supernovae occurring in these diffuse environments. By performing deep, high-cadence difference imaging across the cluster field, we aim to uncover a population of intra-cluster supernovae. The rate and spatial distribution of these events provide a completely independent tracer for the origins and dynamic assembly of the ICM. Additionally, the Coma cluster is known to host a vast population of Ultra-Diffuse Galaxies (UDGs). SNe discovered in these dark-matter-dominated, low-surface-brightness systems will offer critical insights into the star formation history and initial mass functions of UDGs, which remain poorly constrained.

\subsection{Short-timescale Galactic Exotics}
Although the primary targets are extragalactic, the line of sight toward the Coma field intersects the Milky Way halo, capturing numerous galactic stellar sources. Flare stars (e.g., M-dwarfs) and interacting compact binaries frequently exhibit violent outbursts that rise and fall on timescales of mere minutes. Traditional surveys with 30-minute sequential filter cycles completely miss the peak color of these events. Mephisto's 10-minute return time and strictly simultaneous tricolor imaging can faithfully reconstruct the transient spectral energy distribution of ultra-fast galactic flares, distinguishing them from extragalactic fast transients and probing extreme magnetic reconnection physics.

\section{DISCUSSION AND CONCLUSION}
\label{sec:conclusion}

The Mephisto Coma cluster survey offers new opportunities for wide-field time-domain astronomy. By leveraging an innovative mosaic focal-plane design and a highly optimized observational strategy, we have demonstrated the transformative power of simultaneous multi-band, high-cadence imaging. 
During the initial six-month pilot phase, the survey successfully discovered and tracked nine supernovae. 
By targeting the dense and anisotropic environment of the Coma cluster, our survey takes advantage of cosmic large-scale structure to probe transient populations, yielding local transient rates that exceed blind-field expectations and are broadly consistent with our theoretical forecasts that account for stellar mass and star formation rate separately.

Beyond simply inflating discovery statistics, the true scientific legacy of Mephisto lies in its ability to eradicate ``time-skew'' artifacts. Traditional sequential-filter surveys rely on hazardous temporal interpolations to construct spectral energy distributions (SEDs) for fast-evolving transients. In contrast, Mephisto's strictly instantaneous trichromatic colors provide an uncorrupted view of the rapidly changing photosphere in the hours immediately following a stellar explosion. These real-time color diagnostics are indispensable for breaking physical degeneracies in shock-cooling models, probing early circumstellar material (CSM) interactions, and ultimately mapping the final, episodic mass-loss history of massive progenitors. Furthermore, continuous monitoring of such a massive local structure ensures that Mephisto is uniquely primed to provide ultra-rapid, multi-color optical localizations for poorly constrained multi-messenger events, such as gravitational-wave triggers.

Looking ahead, to fully exploit Mephisto's potential in capturing the elusive initial moments of stellar explosions, we plan to deepen the Coma survey by extending the standard exposure time from 45\,s to 180\,s. At a representative local volume boundary of 100\,Mpc ($z \simeq 0.023$ assuming $H_0 = 70\,\mathrm{km\,s^{-1}\,Mpc^{-1}}$, distance modulus $\mu \simeq 35$\,mag), the current 45\,s exposures reach an absolute magnitude limit of $M_{g/r} \lesssim -13$\,mag. By quadrupling the integration time, our sensitivity will improve to $M_{g/r} \lesssim -12$\,mag. This critical enhancement will empower us to detect the extremely faint, ultra-early rising phases of supernovae and precisely track their subtle color evolution immediately following shock breakout. Moreover, pushing the sensitivity to $-12$\,mag will systematically bridge the ``luminosity desert'' between classical novae and sub-luminous supernovae, enabling the discovery of faint, fast-evolving peculiar transients—such as intermediate-luminosity red transients (ILRTs) and luminous red novae (LRNs)—that previously evaded detection. 

In summary, the Mephisto Coma field survey builds on its initial success as a technological pathfinder and provides a valuable platform for studying early-phase supernova physics and exploring short-lived, extreme astronomical phenomena in the local universe.

\section*{Acknowledgements}
Mephisto is developed at and operated by the South-Western Institute for Astronomy Research of Yunnan University (SWIFAR-YNU), funded by the ``Yunnan University Development Plan for World-Class University'' and ``Yunnan University Development Plan for World-Class Astronomy Discipline''. 
This work is supported by the National Key Research and Development Program of China (No. 2024YFA1611603), the National Natural Science Foundation of China (Nos. 12473047,12622307,12303054,12603161), the Yunnan Key Laboratory of Survey Science (No. 202449CE340002), and the Yunnan Fundamental Research Projects (Grant Nos. 202401AU070063, 202501AS070078). 
The authors also acknowledge support from the ``Science \& Technology Champion Project'' (202005AB160002) and from two ``Team Projects'' -- the ``Innovation Team'' (202105AE160021) and the ``Top Team'' (202305AT350002), all funded by the ``Yunnan Revitalization Talent Support Program''.
B.K. acknowledges the support from the ``Special Project for High-End Foreign Experts", Xingdian Funding from Yunnan Province, China. Y.-Z. Cai acknowledges financial support from the SOXS project (PI S. Campana).

\section*{Data Availability}
The data are available at Science Data Bank: \url{https://doi.org/10.57760/sciencedb.014le}.

\bibliographystyle{mnras} 


\bsp    
\label{lastpage}

\end{document}